\documentclass[prd,amsmath,amssymb,aps,floats,amsfonts,notitlepage,superscriptaddress,eqsecnum,nofootinbib,10pt]{revtex4-1}
\usepackage{graphicx}       
\usepackage{dcolumn}        
\usepackage{bm}             
\usepackage{tensor}
\usepackage[utf8]{inputenc}
\usepackage{url}
\usepackage[colorlinks]{hyperref}
\usepackage{color}
\usepackage[usenames,dvipsnames,svgnames,table]{xcolor}
\usepackage{float}
\usepackage{enumerate}
\usepackage[usestackEOL]{stackengine} 
\allowdisplaybreaks[1]

\usepackage{dcolumn}
	\newcolumntype{.}{D{.}{.}{13}}
	\newcolumntype{d}[1]{D{.}{.}{#1}}
\usepackage{longtable}
\usepackage{subfigure}
\usepackage{rotating}
\usepackage[small,compact]{titlesec}
\usepackage{etoolbox}
\makeatletter
\patchcmd{\ttlh@hang}{\parindent\z@}{\parindent\z@\leavevmode}{}{}
\patchcmd{\ttlh@hang}{\noindent}{}{}{}
\makeatother
\usepackage{fancyhdr,amsfonts}
\usepackage{times}
\usepackage{multirow}
\usepackage{sidecap}
\usepackage{eurosym}
\usepackage{setspace}

\definecolor{CiteColor}{rgb}{0,0.5,0}
\hypersetup{citecolor=CiteColor}
\definecolor{RefColor}{rgb}{0.55,0,0}
\hypersetup{linkcolor=RefColor}
\definecolor{darkgreen}{rgb}{0.2,0.7,0.2}

\newcommand{\nn}{\nonumber}

\newcommand{\ord}{\mathcal{O}}

\newcommand{\f}{\frac}

\newcommand{\be}{\begin{equation}}
\newcommand{\ee}{\end{equation}}
\newcommand{\ba}{\begin{eqnarray}}
\newcommand{\ea}{\end{eqnarray}}
\newcommand{\bi}{\begin{itemize}}

\newcommand{\ei}{\end{itemize}}
\newcommand{\bef}{\begin{frame}}
\newcommand{\ef}{\end{frame}}
\newcommand{\es}{\end{small}}

\newcommand\T{\rule{0pt}{2.6ex}}       
\newcommand\B{\rule[-1.2ex]{0pt}{0pt}} 

\begin{document}

\title{Forecasting Gamma-Ray Bursts using Gravitational Waves}

\author{Sarp Akcay}
\affiliation{Theoretisch-Physikalisches Institut, Friedrich-Schiller-Universit{\"a}t Jena, 07743, Jena, Germany}
\affiliation{School of Mathematics \& Statistics, University College Dublin, Belfield, Dublin 4, Ireland}

\begin{abstract}
We explore the intriguing possibility of employing future ground-based gravitational-wave interferometers to detect the inspiral of binary neutron stars sufficiently
early to alert electromagnetic observatories so that a gamma-ray burst (GRB) can be observed in its entirety from its very beginning.
We quantify the ability to predict a GRB by computing the time a binary neutron star (BNS) system takes to inspiral from its moment of detection to its final merger. We define the moment of detection to be the instant at which the interferometer network accumulates a signal-to-noise ratio of 15. 
For our computations, 
we specifically consider BNS systems at luminosity distances of (i) $D\le200\,$Mpc for the three-interferometer Advanced-LIGO-Virgo network of 2020, and (ii) $D \le 1000\,$Mpc for Einstein Telescope's B and C configurations. 
In the case of Advanced LIGO-Virgo we find that we may at best get a few minutes of warning time, thus we expect no forecast of GRBs in the 2020s. 
On the other hand, Einstein Telescope will provide us with advance warning times of more than five hours for $D \le 100\,$Mpc.
Taking one hour as a benchmark advance warning time, we obtain a corresponding range of roughly 600 Mpc for the Einstein Telescope C configuration.
Using current BNS merger event rates within this volume, we show that Einstein C will forecast $\gtrsim \ord(10^2)$ GRBs in the 2030s. 
We reapply our warning-time computation to black hole - neutron star inspirals and find that we expect one to three tidal disruption events 
to be forecast by the same detector.
This article is intended as a pedagogical introduction to gravitational-wave astronomy written at a level accessible to
Ph.D. students, advanced undergraduates, and colleagues in astronomy and/or astrophysics who wish to learn more about the underlying physics.
Though many of our results may be known to the experts, they might nonetheless find this article motivating and exciting.
\end{abstract}
\maketitle

\section{Introduction}\label{Sec:Intro}
The 6.95\footnote{There have thus far been six detections with $5\sigma$ statistical significance and one additional event, LVT151012 \cite{TheLIGOScientific:2016pea}, with $2\sigma \simeq 95\%$ significance hence 6.95 overall.}
\footnote{As of December 2018, the count has been updated to ten including LVT151012 \cite{LIGOScientific:2018mvr}.} 
gravitational-wave merger events detected by the Advanced LIGO-Virgo network have firmly established gravitational-wave astronomy
as an observational science \cite{TheLIGOScientific:2016pea, Abbott:2017gyy, Abbott:2017oio, Abbott:2017vtc, GW170817}. 
Though the first event of 14 September 2015 (GW150914 \cite{GW150914}) will always be the ``poster-child'' of this field,
the icing on the cake was the 17 August 2017 (GW170817 \cite{GW170817}) event involving the inspiral and merger of a binary neutron star system 
promptly followed, 1.7 seconds later, by a gamma-ray burst (GRB170817A) detected by the Fermi Gamma-ray Burst Monitor \cite{Fermi} and by the International Gamma-Ray Astrophysics Laboratory \cite{Svinkin, Savchenko:2017ffs}.
The Advanced-LIGO-Virgo network's initial source localization to within $ 31\,\text{degree}^2$ in the Southern skies
enabled astronomers to locate the electromagnetic counterpart in the galaxy NGC 4993, 
first picked up by the One-Meter, Two Hemisphere team less than 11 hours after the merger \cite{Coulter2017, Coulter:2017wya}.
This subsequently launched a massive campaign of multi-messenger astronomy across the entire electromagnetic (EM) spectrum which is still ongoing
a year after the initial GRB \cite{GBM:2017lvd}. 

GW170817 was initially identified by the LIGO-Hanford interferometer (H1) 
using a template bank of gravitational waveforms computed within the framework of post-Newtonian theory \cite{Buonanno:2009zt, Blanchet_LRR}.
The inspiral swept across the Advanced LIGO-Virgo (ALV) network's bandwidth from approximately $ 30\,$Hz to about $ 2\,$kHz in 57 seconds,
executing nearly $ 3000$ gravitational wave cycles and accumulating a network signal-to-noise ratio of 32.4 with a false-alarm rate of one per $8.0\times 10^4$ years \cite{GW170817}. 
The total mass of the system was inferred to be $2.74^{+0.04}_{-0.01} M_\odot$ with component masses in the range $1.17-1.60 M_\odot$, consistent
with neutron stars (assuming low spin priors, see Sec.~\ref{sec:idealizations}).
Moreover, both the gravitational wave (GW) and EM observations were consistent with a source location in the galaxy in NGC4993 
at a luminosity distance of $\sim\! 40\,$Mpc, and supported the hypothesis that GW170817 resulted from the
inspiral and merger of two neutron stars, causing the prompt short-hard gamma-ray burst GRB170817A. 
This BNS-GRB connection was first proposed in Ref.~\cite{Eichler:1989ve} and the events of 17 August 2017 are
the first astrophysical evidence for it.

A week after GW170817, the ALV network was taken offline for upgrades for the third observation run (O3) 
scheduled to start in early 2019 with roughly twice the sensitivity of O2.
In addition, the Japanese cryogenic interferometer KAGRA (\cite{KAGRA, KAGRA2}) will start its test runs in 2018-2019 and reach its design sensitivity circa 2021-2022 \cite{Akutsu:2017thy, Aasi:2013wya}. By 2025, LIGO-India detector will join the global interferometer network hence improving both the overall sensitivity and the sky-localization capabilities \cite{Aasi:2013wya}.
Around this time, the construction of the first third-generation (3G) ground-based interferometer, Einstein Telescope, should begin in Europe \cite{ET_doc}\footnote{The decision for the funding of Einstein Telescope
is expected to be announced in 2019-2020 with the final technical design report to be submitted in 2023-24, and the commissioning to begin in 2030-31 followed by full operations in 2032-33.}.
Meanwhile, the US is currently considering a mid-2020s update to LIGO called Voyager \cite{LIGO_Voy} and the construction of an ambitious 40-km long 3G interferometer dubbed Cosmic Explorer \cite{CE}.

The science goals of the 3G detectors is very exciting. For example,
Einstein Telescope will be able to (i) distinguish between different neutron star equations of state, (ii) conduct high-precision measurements 
of source parameters to test alternate theories of gravity,
(iii) detect binary black hole populations out to redhifts of $\sim 15$
hence  (a) measure the Hubble parameter, dark matter and dark energy densities, as well as the dark energy equation-of-state parameter;
(b) study the cosmological evolution of the stellar mass black holes (see Sec.~2 of Ref.~\cite{ET_doc} for an excellent summary of the science goals and further details).

The prospect of having extremely sensitive 3G GW detectors begs a simple question: can we detect the inspirals of binary neutron star systems
early enough to witness gamma-ray bursts and the birth of the subsequent kilonovae as they happen? 
In other words, can we forecast GRBs using GW detections?
The short answer is ``yes''. The long answer is still ``yes'', but depends on many factors such as source distance and sky position, detector sensitivity and orientation. One also needs to account for the fact that GRBs are thought to be collimated emissions with narrow outburst angles \cite{Kumar:2014upa}.
As such, most GRBs that occur in the universe are electromagnetically undetectable. Fortunately, this was not the case with GRB170817A, 
but even for a reasonable GRB jet opening angle of $\sim 10^\circ$, we expect to observe at best $\sim 10\%$ of neutron star mergers as GRBs \cite{Patricelli:2016bkt}. 
However, it is possible that some of the near-Earth pointing short hard GRBs may be observed as long-duration GRBs or X-ray flashes thus increasing this percentage \cite{Bucciantini:2011kx}.
On the other hand, the optical emission from the kilonova should be less collimated, thus much more likely
to be detected \cite{Troja:2017nqp}.

Given the sensitivities of interferometers to date, efforts have thus far mostly focused on reducing the latency of EM follow-ups, 
i.e., the lag time between the merger and the alerting of the EM observatories. 
LIGO-Virgo science runs in 2009-2010 yielded a latency of $\sim\!\! 30$ to $60\,$minutes \cite{Abbott:2011ys} (no detections).
In the case of GW170817, the latency was 40\,minutes and $36^{+1.7}\,$seconds \cite{GBM:2017lvd}.
It then took an additional $ 10$ hours to locate the optical transient. 

Here, we are interested  in exploring the predictive power of near-future ground-based GW detectors. 
More specifically, we wish to find out how much early warning future GW detectors can provide us before the merger/GRB.
Ref.~\cite{Cannon:2011vi} approached this from an algorithmic perspective by developing computationally inexpensive methods to detect inspiral
signals in GW data to provide early-warning triggers \emph{before} the merger (also see Ref.~\cite{Knowles:2018hqq} for a very recent improvement to what is currently being
used in the ALV pipeline).
We instead focus on the forecasting capabilities of the ground-based interferometers at their \emph{full design} sensitivity by computing the time interval
between the merger and the instant of detection defined in terms of a certain detection criterion which we describe below.
To this end, we consider the ALV network of 2020s composed of three L-shaped interferometers operating at their design sensitivities
and Einstein Telescope of 2030s made up of V-shaped interferometers.
Ref.~\cite{Nissanke:2012dj} investigated the capabilities of three to five-interferometer detector networks of 2020s in terms of their ranges
to inspiralling BNSs and their sky-localization of these sources. 
More recently, Ref.~\cite{Chan:2018csa} added to this results pertaining to the future networks of 2030s both in terms of source localization and advance
warning prospects.
Our work here is complementary to these articles in the sense that we focus on
the details of how to quantify the advance warning.

We compute the advance warning times via the following procedure: 
(${i}$) We start with a GW source presumed to be an inspiralling BNS at a certain luminosity distance. 
(${ii}$) We determine the frequency at which the source enters a given interferometer's bandwidth. 
(${iii}$) We determine the detection time by computing the frequency at which the network's accumulated signal-to-noise ratio for the BNS inspiral equals 15.
($iv$) We define the advance warning time, $T_\text{AW}$, to be the time interval between the instant of detection and the merger.
We repeat this procedure for BNSs at luminosity distances varying from 50\,Mpc to 1\,Gpc with the smaller values intended for the ALV network and 
the larger ones for Einstein Telescope.
In order to more faithfully represent an interferometer network such as ALV --- consisting of three separate interferometers which have different
orientations and positions --- we employ root-mean-square averages over sky-position and polarization angles,
and assign the interferometers angle-averaged sensitivities.

Our computation for $T_\text{AW}$ is based on the Newtonian evolution of a binary system composed of two point masses in a quasi-circular Keplerian orbit. 
The long-term dynamics is governed by the radiation reaction of 
the GW emission from the system due to the time variance of the quadrupole moment of the binary.
Though the Newtonian (leading-order) treatment gives the 
inspiral time with better than $ 95\%$ accuracy,
we nonetheless supplement it with post-Newtonian corrections up to 3.5 post-Newtonian order (3.5PN) \cite{Blanchet_LRR}.
However, beyond frequencies of $\sim 100\,$Hz the binary enters the strong-field regime where post-Newtonian
approximation become insufficient to provide an accurate depiction of the inspiral.
Instead, the strong-field evolution requires fully
numerical treatment using general relativistic, magnetohydrodynamic codes 
with neutrino transport running on large-scale computing clusters and taking up to millions of CPU hours 
(see Ref.~\cite{Shibata_book, Faber:2012rw} for a review, and Refs.~\cite{Baiotti:2016qnr,Kyutoku:2017voj, Zappa:2017xba, Dietrich:2018upm, Dietrich:2018phi} for the latest developments and references therein).
Nonetheless, despite its simple formulation, our quadrupole radiation driven [post]-Newtonian
evolution is more than sufficient to provide reliable estimations of advance warning times with an error of $\lesssim 1\,$second.
We support this claim in Sec.~\ref{sec:idealizations} with an extensive list of computations which show how much each one of our approximations affects the inspiral time.

%
This article is organized as follows. Sec.~\ref{sec:BNS_inspiral} introduces the formulation for the leading-order evolution of the binary inspiral.
Sec.~\ref{sec:IFO_response} details the interferometer response to GWs based on interferometer topology.
Sec.~\ref{sec:idealizations} lists the various idealizations we employ to simplify our treatment and supplies justification for each.
Sec.~\ref{sec:GW170817} tests our model network and evolution using the parameters of GW170817 and the corresponding observations.
Sec.~\ref{sec:results} contains our main results presented as advance warning times in Tables~\ref{table:LIGO2020} and \ref{table:ET}; 
and as ranges and event rates in Table~\ref{table:horizon}.
In Sec.~\ref{sec:BH_NS}, we recompute the advance warning times in the case of binary black hole - neutron star inspirals to see whether or not 
future detectors can forecast tidal disruption events, for which we summarize our findings in Table~\ref{table:ET_BH_NS}.

Throughout the text, $t$ denotes observer/detector time and $f$ denotes the \emph{GW frequency} of the dominant quadrupole mode in observer/detector frame. 
We employ the $\simeq$ symbol when displaying the numerical values of quantities 
which we usually truncate at four significant digits.
The $\approx$ symbol is reserved for approximations whereas $\sim$ denotes rough, order-of-magnitude equalities,
e.g., $\pi \simeq 3.142$, $\pi \approx 3$, $\pi \sim \ord(1)$. 
Overdots denote time derivatives with respect to detector-frame time, e.g., $\dot{E} =dE/dt$, and $\propto$ denotes proportionalities. 
Unless otherwise noted, we use standard SI units.

\section{Binary neutron star inspirals}\label{sec:BNS_inspiral}
\subsection{Newtonian evolution}
For us, the starting point for the evolution of a binary under gravitational radiation reaction
is the expression for the power emitted in GWs given by 
the celebrated Einstein quadrupole formula \cite{1918SPAW}
%
\be
\dot{E} = \f{G}{5c^5}\langle \dddot{Q}_{ij}\dddot{Q}_{ij} \rangle \label{eq:Edot_quadrupole}\, .
\ee
Here, $Q_{ij}$ is a trace-reversed mass quadrupole moment which
gives the dominant contribution to the power emitted 
by a system with changing mass moments. Details of how to obtain this expression can be found in Ref.~\cite{Maggiore}.
There is no monopole radiation due to conservation of mass and no dipole radiation due to conservation of linear and angular momenta
(cf. Ref.~\cite{Mono}).

Here, we are specifically interested in the motion of two point masses in a circular orbit around a common center of mass.
For a binary in a circular orbit with Keplerian angular frequency $\Omega$, separation $r$, and masses $m_1 \le m_2$, we have \cite{Maggiore}
\be
\dot{E} = \f{32}{5}\f{G\mu^2}{c^5}\, r^4 \Omega^6\label{eq:Edot},
\ee
where $\mu= m_1 m_2/M$ is the reduced mass and $M=m_1+m_2$ is the total mass. 
We justify our simplification of the motion to circular orbits in Sec.~\ref{sec:idealizations}.

An important quantity in GW astronomy is the chirp mass of the binary given by
\be
M_c = \mu^{3/5} M^{2/5} = \f{(m_1 m_2)^{3/5}}{(m_1+m_2)^{1/5}} \label{eq:chirp_mass}\, ,
\ee
which turns Eq.~(\ref{eq:Edot}) into
\be
\dot{E} = \f{32}{5}\f{c^5}{G} \left(\f{G M_c\, \omega}{2c^3}\right)^{10/3}= \f{32}{5}\f{c^5}{G} \nu^2 x^5, \label{eq:Edot2}
\ee
where $\omega=2\Omega$ is the angular frequency of the quadrupolar GWs, $\nu\equiv \mu/M$ is the symmetric mass ratio and $x = ({G M\, \Omega}/{c^3})^{2/3} $ is the dimensionless inverse separation parameter ubiquitous 
in post- Newtonian theory. 
As $x\propto v^2/c^2$, it is used to keep track of post-Newtonian orders and we can see from Eq.~(\ref{eq:Edot2})
that, at the leading order, power emitted in GWs scales as $x^5$.

The emission of power in GWs decreases the total energy of the binary given by
\be
E_b = -\frac{G m_1 m_2}{2r}\,.  \label{eq:binding_energy1}
\ee
$\dot{E}_b \propto \dot{r}/r^2$, so $\dot{E}_b < 0$ implies $\dot{r}<0$, i.e, the orbital separation decreases
making the orbit more bound in time.
Using Kepler's third law, $ \Omega^2 = G M r^{-3}$, and Eq.~(\ref{eq:chirp_mass}) we obtain
\be
E_b = -\left(\f{G^2 M_c^5 \omega^2}{32}\right)^{1/3}\, . \label{eq:binding_energy2}
\ee
Energy conservation dictates that $\dot{E}=-\dot{E}_b$. Using the chain rule to determine $\dot{E}_b$ from Eq.~(\ref{eq:binding_energy2})
and setting the resulting expression equal to Eq.~(\ref{eq:Edot2}) yields $\dot{\omega}$ as a function of $\omega$ and various constants.
Translating this expression using $\omega=2\pi f$ we arrive at the standard expression for the GW frequency evolution
\be
\dot{f} = \f{96}{5}\pi^{8/3} \f{(G M_c)}{c^5}^{5/3}\, f^{11/3} \label{eq:fdot}
\ee
This can be integrated straightforwardly after defining a new time variable $\tau\equiv t_\text{coal}-t$ that equals zero when the neutron stars coalesce.
Using $\tau$ we can now define the inspiral time, i.e, time left to coalesence at a given GW frequency $f$
\be
\tau_\text{insp}(f) = \f{5}{256\pi}\f{c^5}{(\pi G M_c)^{5/3}} \,f^{-8/3}\label{eq:tau_insp}\, ,
\ee
%
%
%
which we can rewrite as follows
\be
\tau_\text{insp}(f) \simeq 16.72\,\text{minutes} \, \left(\f{1.219 M_\odot}{M_c}\right)^{5/3}\,\left(\f{10\,\text{Hz}}{f}\right)^{8/3}, \label{eq:tau_insp2}
\ee
where $1.219 M_\odot$ is the chirp mass corresponding to $m_1=m_2=1.4 M_\odot$. This is the canonical value for the 
mass of a neutron star due to the Chandrasekhar limit \cite{Chandra}.

We can also compute the number of GW cycles over the course of an inspiral. Given that the orbital period $T_K(t)$ varies over a time scale much larger than itself
we can write
\be
\mathcal{N}_\text{cyc}(f) = \int_f^{f_\text{max}} \f{f'}{\dot{f'}}\, df' =\f{1}{32\pi} \f{c^5}{(\pi G M_c)^{5/3}}\left[ f^{-5/3}-f_\text{max}^{-5/3}\right] \label{eq:Ncyc1}\, .
\ee
We now introduce a cut-off for the inspiral imposed by the frequency of the innermost stable circular orbit (ISCO) in Schwarzschild spacetime
\be
\Omega_\text{ISCO} = \f{c^3}{6^{3/2} G M} \label{eq:Sch_f_isco}.
\ee
We will see in Sec.~\ref{sec:idealizations} that $f_\text{ISCO} \sim \ord(10^3)\,$Hz for a system with $M =2.8 M_\odot$. 
On the other hand, the frequencies of interest for our advance warning time estimations are $\lesssim 10\,$Hz
meaning that $f^{-5/3} \gg f_\text{ISCO}^{-5/3}$. Therefore, at the leading (Newtonian) order, the number of GW cycles can be approximated by
\be
\mathcal{N}_\text{cyc}(f) \approx \f{1}{32\pi} \f{c^5}{(\pi G M_c)^{5/3}}\, f^{-5/3} \,  \label{eq:Ncyc2}
\ee
%
%
which can be rewritten as
\be
\mathcal{N}_\text{cyc}(f)\approx 1.605\times 10^4 \,\left(\f{1.219 M_\odot}{M_c}\right)^{5/3}\, \left(\f{10\,\text{Hz}}{f} \right)^{5/3}\label{eq:Ncyc3}\, .
\ee
This expression is quite telling: if a ground-based interferometer picks up an
inspiralling BNS at $f=10\,$Hz then there will be $\gtrsim \ord(10^4)$ GW cycles in the detector's data stream until the merger.

Another useful relation is how the inspiral time scales with respect to the orbital radius corresponding to the observed GW frequency.
From Kepler's third law, we immediately have $\dot{r}/r= -2\dot{\Omega}/(3\Omega)=-2\dot{f}/(3f)$, which, via Eq.~(\ref{eq:tau_insp}),
yields $\dot{r}/r= -1/(4\tau)$ which integrates to
\be
r(\tau)=r_i \left(\f{\tau}{\tau_i}\right)^{1/4}\label{eq:r_of_tau},
\ee
where $r_i, \tau_i=t_\text{coal}-t_i$ are the initial radius and time that the BNS
is ``picked up'' by a detector. Solving $\tau(f_i)=\tau_i$ for $f_i=\Omega_i/\pi$ using Eq.~(\ref{eq:tau_insp}) and rewriting $\Omega_i$ as a function of $r_i$ via
Kepler's third law gives us
\be
\tau_i = \f{5}{256}\, \f{c^5 r_i^4}{G^3 M^2 \mu}\simeq 1.024\,\text{hr} \left(\f{r_i}{1000\,\text{km}}\right)^4 \left(\f{2.8M_\odot}{M}\right)^2\,\left(\f{0.7M_\odot}{\mu}\right)
\label{eq:tau_of_r},
\ee
where the seemingly arbitrary $0.7M_\odot$ is the value of $\mu$ for $m_1=m_2=1.4M_\odot$.

Let us now turn our attention to the GWs generated by the inspiral.
These are tensorial perturbations propagating at the speed of light in flat (Minkowski) spacetime.
In the so-called wave zone, defined by the condition $ \lambda= c/f \ll D$, where $\lambda$ is the GW wavelength and $D$ is the distance to the GW source, 
these tensor modes satisfy the flat spacetime sourceless wave equation. 
%
It can further be shown that there exist only two radiative degrees of freedom (cf. Sec.~1.2 of Ref.~\cite{Maggiore}) which physically manifest themselves as two independent polarization amplitudes $h_+$ and $h_\times$ of the GWs. 
This fact is most easily demonstrated in the so-called transverse-traceless gauge (cf. Sec.~2.1 of Ref.~\cite{Flanagan:2005yc}),
 but is a gauge-invariant consequence of linearized vacuum solutions. 

The two polarization states generated by a binary in a circular orbit read \cite{Maggiore}
\begin{align}
 h_+(t) &= h_c(t)\, \left(\f{1+\cos^2\iota}{2}\right)\, \cos[\Phi_\text{N}(t)],\label{eq:hplus_TD}\\
 h_\times(t) & =h_c(t)\,\cos\iota \sin[\Phi_\text{N}(t)]\label{eq:hcross_TD},
\end{align}
where
\be
h_c(t) =\f{4}{D}\left(\f{G M_c}{c^2}\right)^{5/3}\, \left(\f{\pi f(t)}{c}\right)^{2/3} \label{eq:h_c}
\ee
is the characteristic strain and
$\iota=\cos^{-1}(\mathbf{\hat{n}}\cdot{\mathbf{\hat{L}}}) $ is the inclination angle between the line of sight unit vector $\mathbf{\hat{n}} $ and the orbital angular momentum unit vector $\mathbf{\hat{L}}$. 
For an orbit seen edge on, $\iota = \pi/2$ giving $h_\times=0$, i.e,  a purely plus-polarized wave.
When $\iota=0$ the orbit is seen face-on and we have a circularly-polarized wave: $\langle h_+ \rangle = \langle h_\times\rangle$ where $\langle \ldots \rangle$
denote time-averages over one orbit. Note that as $f$ increases in time, so does $h(t)$, hence the characteristic ``chirping'' of GW signals.

$\Phi_\text{N}(t)$ in Eqs.~(\ref{eq:hplus_TD}, \ref{eq:hcross_TD}) is the phase of the GWs given by
\begin{align}
\Phi_\text{N}(t) &=\int_{t_i}^t dt' \omega(t') \nn\\
&= -2\left(\f{5 G M_c}{c^3}\right)^{-5/8} (t_\text{coal}-t)^{5/8}+ \Phi_0 \, ,\label{eq:Phase}
\end{align}
where $\Phi_0 \equiv \Phi(t=t_\text{coal})$ is an integration constant. The subscript N denotes the Newtonian (leading-order) contribution. 
Higher-order contributions can be added in terms of a post-Newtonian series, which make up to $\lesssim 2\%$ of the total phase.
Although we will include contributions up to and including 3.5PN to obtain the results of Sec.~\ref{sec:results}, 
we will not show these expressions here, but they can be found, e.g., in Ref.~\cite{Blanchet_LRR}.

The nomenclature ``plus'' ($+$) and ``cross'' ($\times$) follows from the effects that passing GWs have on the plane transverse to their direction of propagation.
If we were to conceive of a ``toy'' detector made up of a circularly arranged test-masses lying in the $x$-$y$ plane, a GW propagating along the $z$ direction would stretch/compress this circular arrangement in a $+$ and $\times$ pattern 
with a $90^\circ$ phase difference between the two states.
This is the manifestation of the tidal strain of the GWs on the tranverse plane.
Next, let us briefly explore how an actual detector, i.e., an interferometer, responds to passing GWs.

\section{Interferometer response to gravitational waves}\label{sec:IFO_response}
\subsection{LIGO-Virgo configuration: L-shaped topology}\label{sec:LIGO_topo}
We start by first considering the L-shaped interferometer topology dating back to Michelson \cite{Michelson:1887zz}.
Although GWs are described by propagating tensor modes, an interferometer (IFO) can only measure a scalar quantity
known as the response function (or GW strain) which is a linear combination of the polarizations given by
\be
h(t) = F_+(\theta,\phi,\psi) \,h_+(t)+F_\times(\theta,\phi,\psi)\, h_\times(t) , \label{eq:detector_strain}
\ee
where
\begin{align}
 F_+ &=\f{1}{2}\left(1+\cos^2\theta\right)\cos2\phi\cos 2\psi-\cos\theta \sin2\phi \sin2\psi \label{eq:F_plus},\\
 F_\times &=\f{1}{2}\left(1+\cos^2\theta\right)\cos2\phi\sin2\psi+\cos\theta \sin2\phi \cos2\psi \label{eq:F_cross}
\end{align}
%
are the antenna pattern functions of the detectors; $\theta,\phi$ are the sky coordinates; and $\psi$ is
the angle of orientation of a detector's frame with respect to the source frame. 
%
We can not apriori know $\theta, \phi,\psi$, however, we can extract their values with some limited precision over the course of an inspiral.
The longer the inspiral stays in the detector bandwidth, the better we can estimate these angles.
In the the case of GW170817 which swept the ALV network over roughly 3000 GW cycles, 
the initial rapid source sky localization was 31\,deg$^2$ (later reduced to 28\,deg$^2$) using data from three IFOs \cite{GW170817}. 
For our estimations, we employ suitable averages of these angles as explained below.
For details on how to obtain Eqs.~(\ref{eq:F_plus}, \ref{eq:F_cross}) see, e.g., Sec.~4.2.1 of Ref.~\cite{SchutzLRR}.

To compute advance warning times we need to know how ``loud'' neutron star binaries become as they sweep across an IFO's detection bandwidth.
This is quantified in terms of the GW signal-to-noise ratio (SNR) accumulated in a detector's bandwidth during the inspiral.
The SNR is computed in the frequency domain so we first Fourier-transform the time-domain strains of Eqs.~(\ref{eq:hplus_TD}) and (\ref{eq:hcross_TD}) to obtain
\begin{align}
 \tilde{h}_+(f) &= A \, \f{c}{D}\, \left(\f{G M_c}{c^3}\right)^{5/6}\, e^{i \Psi_+(f)}\, \f{1}{f^{7/6}}\,  \f{1+\cos^2\iota}{2}  ,\label{eq:hplus_FD}\\
 \tilde{h}_\times(f) & = A \, \f{c}{D}\, \left(\f{G M_c}{c^3}\right)^{5/6}\, e^{i \Psi_\times(f)}\, \f{1}{f^{7/6}}\, \cos\iota \label{eq:hcross_FD},
\end{align}
where $A= \pi^{-2/3} \sqrt{5/24}$.
Explicit expressions for $\Psi_{+,\times}$ show that they are out of phase by $\pi/2$
(the details of the Fourier transformation can be found in Sec.~4.5 of Ref.~\cite{Maggiore} or in Ref.~\cite{Moore:2014lga}).
Accordingly, the Fourier transform of the detector strain (\ref{eq:detector_strain}) is given by
\be
\tilde{h}(f) = A\, \f{c}{D}\left(\f{G M_c}{c^3}\right)^{5/6} f^{-7/6}\, e^{i\Psi_\text{N}}\, Q(\theta,\phi,\psi,\iota), \label{eq:strain_FD}
\ee
where 
\be
Q(\theta,\phi,\psi,\iota) = F_+(\theta,\phi,\psi)\f{1+\cos^2\iota}{2}  + i F_\times(\theta,\phi,\psi) \cos\iota \label{eq:Q}
\ee
is the quality factor.
The $i$ in front of $F_\times$ in $Q(\theta,\phi,\psi,\iota)$ is due to the $\pi/2$ phase difference between $+$ and $\times$ modes.
$\Psi_\text{N}$ is the leading-order contribution to the frequency-domain phase
\be
\Psi_\text{N}=\Psi_+(f)|_\text{N} = 2\pi f (t_\text{coal}+D/c)-\Phi_0-\f{\pi}{4}+ \f{3}{128}\left(\pi f\,\f{G M_c}{c^3}\right)^{-5/3} \label{eq:phase_FD}.
\ee
Higher-order contributions to $\Psi_{+,\times}$ up to and including 3.5PN can be found in Ref.~\cite{SchutzLRR}.

The detection criterion we employ here is quantified in terms of the optimal SNR defined by
\be
\rho=\left[ \int_0^\infty d\ln f\, \f{|2\tilde{h}(f)\sqrt{f}|^2}{S_n(f)}\right]^{1/2} \label{eq:SNR},
\ee
where $\sqrt{S_n(f)}$ is the {\it amplitude spectral density} (ASD) of the detector (also called spectral strain sensitivity, or simply, detector noise).
It is usually this quantity that is shown in a typical interferometer strain sensitivity plot with the characteristic strain, $|2\tilde{h}(f)\sqrt{f}|$, 
due to various GW sources overlaid for comparison. 
%
We show such plots in Secs.~\ref{sec:GW170817} and \ref{sec:results}.

Substituting Eq.~(\ref{eq:strain_FD}) into Eq.~(\ref{eq:SNR}) yields
\be
\rho^2 = \f{5}{6}\, \pi^{-4/3} \f{c^2}{D^2}\left(\f{G M_c}{c^3}\right)^{5/3}|Q(\theta,\phi,\psi,\iota)|^2 \int_0^{f_\text{ISCO}} d f\, \f{f^{-7/3}}{S_n(f)}\, \label{eq:SNR_v2},
\ee
where we truncate the integral at the cut-off frequency of the inspiral $f_\text{ISCO}=\Omega_\text{ISCO}/\pi$.

It remains to compute $|Q(\theta,\phi,\psi,\iota)|^2$. 
Since we can not a priori know the direction of the source, it is appropriate to average over the angles $\theta,\phi,\psi$, and $\iota$ which we compute
%
%
via the following RMS-averaging integral
\be
\langle F^2_{+}\rangle \equiv \f{1}{2\pi}\int_0^{2\pi}d\psi \,\f{1}{4\pi}\int_0^{2\pi} d\phi\,\int_0^\pi d\theta \sin\theta\ F^2_{+}(\theta,\phi,\psi)=\f{1}{5}\label{eq:ang_av}
\ee
and likewise $\langle F^2_{\times}\rangle=1/5 $.
Finally, we average over the orbital inclination angle
\be
\f{1}{2} \int_{-1}^1 d(\cos\iota)\left[\left(\f{1+\cos^2\iota}{2}\right)^2+\cos^2\iota\right] = \f{4}{5} .\label{eq:inc_ang_av}
\ee
Thus we have that
\be
\langle |Q(\theta,\phi,\psi,\iota)|^2 \rangle_{\theta,\phi,\psi,\iota} = \f{4}{25}\label{eq:RMS_th_ph_ps_iota}.
\ee
As angles have zero averages in general, we can think of the RMS average as the standard deviation ($\sigma =2/5$). 
%
%
%
%
%

In a network of $N$ interferometers, the total SNR is defined as follows \cite{Finn:2000hj}
\be
\rho_\text{tot} = \left[{\sum_{i=1}^N \rho^2_i}\right]^{1/2}\, , \label{eq:SNR_total}
\ee
where Eq.~(\ref{eq:SNR_v2}) is employed to compute $\rho^2_i$ of individual interferometers each with a specific quality factor $Q_i$.
Here, instead of working with specific quality factors of the LIGO and the Virgo interferometers, we consider
a \emph{model network} of three IFOs ($N=3$), where one operates at the RMS average; thus it has $|Q|=2/5$. 
The second IFO is better oriented/positioned so we assign it a quality
factor one $\sigma$ above the average: $|Q|=2/5+\sigma=4/5$.
It remains to assign a quality factor to the third IFO. We wish to make this operate below the RMS average, 
but it would be pointless to simply set $Q=0$. We therefore pick the values $\iota=0$ (i.e, $|h_+|=|h_\times|$)
and $\psi=0$ for simplicity 
then introduce the following $\{\theta,\phi\}$-only averages
\be
 \langle F_+^2\rangle_{\theta,\phi} = \frac{7}{30},\qquad \langle F_\times^2\rangle_{\theta,\phi} = \frac{1}{6}\, 
\ee
yielding
\be
\langle |Q(\theta,\phi,\psi=0,\iota=0)|^2 \rangle_{\theta,\phi} = \f{2}{5} \label{eq:RMS_th_phi}.
\ee
Using this value, we assign our below-average third IFO 
$|Q|=4/5-\sqrt{2/5}\simeq 0.1675$.
This may seem somewhat ad hoc, 
but it is our way of representing the sub-RMS-average response of the third IFO due to its sub-optimal configuration. 
One may very well use another scheme to generate this below-average response.
Given that $\rho_\text{tot}$ is a sum of quadratures the lowest SNR contributes a small percentage to this [cf. Eq.~(\ref{eq:SNR_total})].

To summarize, we introduced a model for a three-IFO network in which each IFO has identical sensitivity,
but varying response to GWs due to differing IFO orientations and source polarization and sky position. This response is encoded in the quality factor $|Q_i|$.
%
%
%
Accordingly, we now introduce the norm of the frequency-domain strain (\ref{eq:strain_FD}) 
\be
\tilde{H}_i(f)\equiv \tilde{A}_i\, h_0\, f^{-7/6} \label{eq:Hi_s}
\ee
for each IFO labelled by $i$, where
\begin{align}
h_0 &= \f{c}{D}\left(\f{G M_c}{c^3}\right)^{5/6} \label{eq:h0}\, 
\end{align}
and $\tilde{A}_i\equiv A |Q_i|$ 
are given in Table~\ref{table:network_params}. 
The SNR for each interferometer can now be written as
\begin{align}
\rho_i &= 2\tilde{A}_i\, h_0 \left[\int_0^{f_\text{ISCO}} df\, \f{f^{-7/3}}{S_\text{n}(f)} \right]^{1/2}\label{eq:SNR_i}.
\end{align}
We arbitrarily make the following assignments for our network: $h=1, m=2, l=3$ representing high, medium, low signal strain
at IFO$_i$, respectively. 
We summarize our network in the following table.

\begin{table}[h]
\centering
\begin{tabular}{ccccc}
\hline\hline
 IFO$_i$  
 & $|Q_i|$ & $\tilde{A}_i$ & $\tilde{H}_i(f)$ & IFO response\T\B \\
 \hline
 1 &  $\tfrac{4}{5} $&  $\tfrac{4}{5}A$ & $\tfrac{4}{5}A h_0\, f^{-7/6}$ & high ($h$) \T\\
 2 &  $\tfrac{2}{5} $ & $\tfrac{2}{5}A$ & $ \tfrac{2}{5}A h_0 \, f^{-7/6}$ & medium ($m$)\B\T \\
 3 &  $\quad \tfrac{4}{5}-\sqrt{\tfrac{2}{5}} \quad $
 & $\quad\left(\tfrac{4}{5}-\sqrt{\tfrac{2}{5}}\right)A \quad$ & $\quad \left(\tfrac{4}{5}-\sqrt{\tfrac{2}{5}}\right)A h_0 \, f^{-7/6}\quad$& low ($l$) \B \\
 \hline\hline
\end{tabular}
\caption{
Our model for the Advanced-LIGO-Virgo network ca. 2020. 
Column two shows the quality factor $Q_i$ for each IFO and column three the corresponding amplitude scalar $\tilde{A}_i$ introduced
in Eq.~(\ref{eq:Hi_s}).
Column four lists the magnitude of the frequency-domain strain at each IFO via Eq.~(\ref{eq:Hi_s}).
The last column states each IFO's response, e.g., high means high sensitivity, thus high SNR.
Recall $A= \pi^{-2/3} \sqrt{5/24}$. Maximum value for $|Q_i|$ is 1 and minimum is 0.
}\label{table:network_params}
\end{table}
We can now write the total SNR for our three L-IFO network
\begin{align}
\rho_\text{tot} &= \sqrt{\rho^2_h+\rho^2_m+\rho^2_l} \label{eq:SNR_total_3IFOs} 
\end{align}
with $\rho_h, \rho_m,\rho_l$ given by Eq.~(\ref{eq:SNR_i}) and Table~\ref{table:network_params}.
%

Let us conclude this subsection with a brief discussion on 
an alternate approach to construct a model for the near-future ALV network.
This involves selecting one IFO as the ``base'' detector, e.g., LIGO Livingstone, then running Monte-Carlo simulations over the four random angles $\{\theta,\phi,\psi,\iota\}$. 
We would subsequently compute the quality factor of the base IFO using the
simulation results for the four angles.
We can then obtain the quality factors of the remaining two IFOs 
--- in this particular case, LIGO Hanford and Virgo --- via their positions and orientations with respect to the base IFO. Given that the four angles
are completely random, there should be no preferred region in the angle parameter space, thus the result of the Monte-Carlo runs should more or less agree with the RMS average. Correspondingly, we would expect one of the other two IFOs to have somewhat higher sensitivity, and the other one to have lower sensitivity.
This network can be further expanded using exact positions of LIGO India
and KAGRA detectors. The expansion of the detector network from three to five IFOs will offer significant improvements in source localization.
However, given that the total network SNR is a sum of quadratures, the 
expanded network will not accumulate significantly higher SNR than
our model ALV network. As such our model network should serve as a good proxy for the advance warning time computations.

\subsection{Einstein Telescope configuration: triangular topology}\label{sec:ET_topo}
Current design of Einstein Telescope (ET) is based on a [equilateral] triangular configuration with 10\,km armlengths \cite{ET_doc}.
Within this equilateral triangle, ET will consist of three V-shaped cryogenic interferometers housed underground to
significantly reduce seismic and gravity-gradient noises \cite{ET_doc}.
This closed topology will allow the detector to form a null stream completely devoid of a signal \cite{Sathyaprakash:2012jk},
which can be used to rule out spurious events \cite{Wen:2005ui}.
Additionally, as we show below, the response function and the quality factor of the triangular configuration are independent of the azimuthal angle $\phi$
hence ET will have no blind spots \cite{Regimbau:2012ir}.
However, the $60^\circ$ arm separation reduces the strain sensitivity of a V-shaped interferometer by a factor of $\sin 60^\circ=\sqrt{3}/2$ compared
to an L-shaped interferometer with the same armlength. Thus, for a single V-IFO the antenna patterns of Eqs.~(\ref{eq:F_plus}, \ref{eq:F_cross}) become
\be
F^1_+(\theta,\phi,\psi)= \f{\sqrt{3}}{2} F_+(\theta,\phi,\psi), \qquad F^1_\times(\theta,\phi,\psi)= \f{\sqrt{3}}{2} F_\times(\theta,\phi,\psi)\label{eq:F1plus_F1cross}\, ,
\ee
where the superscript $1$ labels one of the three V's of the triangle.
Since the other two V-IFOs lie in the same plane as the first one, their antenna patterns can be obtained simply by rotating in the azimuthal
direction by $120^\circ, 240^\circ$, respectively
\begin{align}
 F^2_{+,\times}(\theta,\phi,\psi) &= F^1_{+,\times}(\theta,\phi+2\pi/3,\psi), \label{eq:F2plus_F2cross}\\
 F^3_{+,\times}(\theta,\phi,\psi) &= F^1_{+,\times}(\theta,\phi-2\pi/3,\psi) \label{eq:F3plus_F3cross}.
\end{align}
The invidual antenna responses combine to cancel the $\phi$ and $\psi$ dependence of the total power pattern
\be
F^2_\bigtriangleup \equiv \sum_{A=1}^3 \left(F^A_+\right)^2+\left(F^A_\times\right)^2 = \f{9}{32}\left(1+6\cos^2\theta + \cos^4\theta\right). \label{eq:ET_power_pattern}
\ee
From this, we immediately have that the RMS average is given by $\langle F^2_\bigtriangleup\rangle^{1/2} =3/\sqrt{10}$ and the minium value is $F_\bigtriangleup(\pi/2)=3/\sqrt{32}$;
thus a triangular detector made-up of three V-IFOs has all-sky coverage.

For the total SNR accumulated in a triangular detector we have
\begin{align}
\rho^2_{\text{tot},\bigtriangleup} & = \sum_{A=1}^3 \rho_A^2 \nn \\ 
 & =  4 A^2 h_0^2\,|Q(\theta,\phi,\psi,\iota)|_\bigtriangleup^2 \int_0^{f_\text{ISCO}} d f\, \f{f^{-7/3}}{S_n(f)} \, ,\label{eq:ET_SNRsq}
\end{align}
where $h_0$ is given by Eq.~(\ref{eq:h0}) and $A=\pi^{-2/3}\sqrt{5/24}$ as before and
\be
|Q(\theta,\phi,\psi,\iota)|_\bigtriangleup^2 = \sum_{A=1}^3 \left(F^A_+\right)^2\f{1+\cos^2\iota}{2}  + \left(F^A_\times\right)^2 \cos\iota \label{eq:Q_sq_ET}
\ee
which is independent of $\phi$ \cite{Regimbau:2012ir}. We once again take the RMS average of this quantity:
%
\be
\bar{Q}^2_\bigtriangleup\equiv\langle |Q(\theta,\phi,\psi,\iota)|_\bigtriangleup^2 \rangle_{\theta,\phi,\psi,\iota} = \f{9}{25}\label{eq:RMS_ET}
\ee
which is larger than a single L-shaped IFO's RMS average of 4/25 given in Eq.~(\ref{eq:RMS_th_ph_ps_iota}).

As we can not a priori know the source sky position or polarization we model the response of ET using the RMS-averaged SNR 
\be
\rho_{\text{tot},\bigtriangleup} = \f{6}{5}A\, h_0 \left[\int_0^{f_\text{ISCO}} d f\, \f{f^{-7/3}}{S_n(f)}\right]^{1/2} \label{eq:ET_SNR},
\ee
Accordingly, we define an RMS-averaged frequency-domain strain
\begin{align}
\tilde{H}_\bigtriangleup(f)&\equiv A\,  \bar{Q}_\bigtriangleup\, h_0 f^{-7/6} \nn \\
   & = \f{1}{2}\sqrt{\f{3}{10}}\,{\pi^{-2/3}}\, \f{c}{D}\left(\f{G M_c}{c^3}\right)^{5/6} f^{-7/6}. \label{eq:H_delta}
\end{align}
This is ET's sky-averaged response to a binary inspiral at luminosity distance $D$ and chirp mass $M_c$.
Thus, $\tilde{H}_\bigtriangleup(f)$ is like the $\tilde{H}_i(f)$ of Table~\ref{table:network_params},
but different in that  $\tilde{H}_\bigtriangleup(f)$ represents the total response of the entire three V-IFO network whereas $\tilde{H}_i(f)$
is the response of the $i^\text{th}$ L-IFO.

In Sec.~\ref{sec:results} we will consider BNS systems inspiralling at varying luminosity distances as typical sources for the ALV and ET networks.
To compute the resulting SNRs, we will slightly modify the integrals in Eqs.~(\ref{eq:SNR_i}, \ref{eq:ET_SNR}) and 
use design sensitivities of A-LIGO and ET for $\sqrt{S_n(f)}$.
Before we present our results, we list the various idealizations that we have adopted to simplify our computations.
For each idealization we present an estimated error. 
Our overall conclusion is that our simplifications do not significantly change our estimations of advance warning times.
%
%
%
%
\section{Simplifications and idealizations} 
\label{sec:idealizations}
\begin{enumerate}
 \item {\it Neglecting strong-field gravity.}
As our aim here is to provide a good order-of-magnitude estimation for the advance warning times, we model the neutron stars as point masses 
all the way to the merger. The point-particle treatment is severely inadequate for the strong-field evolution of the binary, but these systems spend only the last few seconds of the inspiral in this regime, e.g., the inspiral time from $f=100\,$Hz to the merger is
$\tau_\text{insp}(f=100\,\text{Hz})\simeq 2.161\,$seconds according to Eq.~(\ref{eq:tau_insp}).
And even at $f=100\,$Hz, a $m_1=m_2=1.4M_\odot$ neutron-star binary has separation $r\approx 155\,$km, which translates to a dimensionless strength of gravity of $\approx 2.6\%$.
This percentage is mostly accounted for by the post-Newtonian
corrections to our Newtonian (leading order) evolution. 
Here, we will go up to 3.5pN for our
advance warning time computations.
Therefore, given that strong-field effects may at most change
$T_\text{AW}$ by a second, we can neglect them for our purposes,
but let us add that a faithful evaluation of BNS systems above $f\gtrsim 100\,$Hz is a very active and important branch of general relativity and gravitational-wave astronomy. 
%
\item {\it Terminating the inspiral at the Schwarzschild ISCO.}
We artificially end the inspirals at the Schwarzschild ISCO of the BNS with coordinate radius $r_\text{ISCO} = 6GM/c^2$ and angular frequency $\Omega_\text{ISCO}$ 
given by Eq.~(\ref{eq:Sch_f_isco}) which translates to the following dominant-mode (quadrupole) GW frequency
\be
f_\text{ISCO} = \f{c^3}{6^{3/2}\pi G M} \simeq 1571 \left(\f{2.8M_\odot}{M}\right)\text{Hz} \label{eq:f_isco}.
\ee
In the strong-field, highly-dynamical spacetime of the last few orbits, the
concept of an ISCO becomes fuzzy due to strong dissipative effects. However, an ISCO can nonetheless be dynamically determined
with frequency possibly up to 1.7\,kHz \cite{Marronetti:2003hx}, 
but there is no simple expression for it akin to Eq.~(\ref{eq:f_isco}) and its extraction from numerical simulations is a rather involved procedure.
Therefore, we employ Eq.~(\ref{eq:f_isco}) as our cut-off for
inspirals throughout this work.
%
\item {\it Setting} $m_1=m_2=1.4 M_\odot$.
Let us briefly explore what the cost of setting $m_1=m_2=1.4 M_\odot$ is for our calculations. 
Including the error bars, the known range for neutron star masses runs from $\approx 0.5 M_\odot$ to $ \approx 3 M_\odot$ with typical values between $\approx 1 M_\odot$ and $\approx 2 M_\odot$ \cite{Ozel_Freire} (one exception being J1748-2021B with mass $2.74^{+0.21}_{-0.21} M_\odot $ \cite{Freire_et_al_2007}, but this measurement is not secure \cite{Lattimer:2012nd}\footnote{We thank Dr. Alberto Sesana of University of Birmingham for pointing this out.}).
Of these measured masses, a subset comprised of double neutron stars, such as the famed Hulse-Taylor binary, have the smallest error bars with masses
ranging from $\approx 1.2 M_\odot$ to $\approx 1.5 M_\odot$. 
On the other hand, the masses involved in GW170817 are inferred to be between $1.17 M_\odot$ and $1.60 M_\odot$ \cite{GW170817}.
Therefore, let us restrict the masses to lie between $1.2 M_\odot$ and $1.6 M_\odot$, then focus on 
the quantities of interest for us, which are 
the inspiral time ($\tau_\text{insp}$), the frequency domain strain ($\tilde{H}$), and
the frequency of the innermost stable circular orbit ($f_\text{ISCO}$) given by Eq.~(\ref{eq:f_isco}) above.
We summarize how these scale in terms $M_c, M$, thus $m_1, m_2$, in Table \ref{tbl:scalings} below.
\begin{center}
\begin{table}[ht]
\begin{tabular}{lcccc}
\hline\hline
 Quantity\hspace{5mm} & Standard Scaling & \hspace{3mm}Scaling in terms of $m_1,m_2$ & &\hspace{0.5cm} Min $\le$ Value$_{1.4M_\odot} \le$ Max\T\B\\
\hline
$\tau_\text{insp}$ & $M_c^{-5/3} $ & $(m_1 m_2)^{-1}(m_1+m_2)^{-1} $&\quad& $0.335\le 0.5\le 0.794 $\T \\
$\tilde{H}$ & $M_c^{5/6} $ & $(m_1 m_2)(m_1+m_2)^{3/2} $&\quad& $ 0.7834\le 0.891 \le 0.996 $ \T\B\\
$f_\text{ISCO}$ & $M^{-1} $ & $(m_1+m_2)^{-1} $&\quad& $0.4375\le 0.5\le 0.583 $\B\\
\hline\hline
\end{tabular}
\caption{The scaling of the physical quantities of interest in this article in terms the binary masses $m_1, m_2$ restricted to lie between $1.2M_\odot$ and $1.6 M_\odot$. 
Value$_{1.4M_\odot}$ represents the value of the quantities in column three evaluated at $m_1=m_2=1\, \text{mass unit}$ corresponding to $1.4 M_\odot$.
}
\label{tbl:scalings}
\end{table}
\end{center}
\vspace{-11mm}
The ranges in column four show that $\tilde{H}$ changes roughly by $\pm 10\%$ across the $1.2M_\odot-1.6M_\odot$ mass range whereas the variation
in $f_\text{ISCO}$ should not matter as it affects the advance warning times by less than a second (see above).
The inspiral time seems to be more sensitive to our chosen mass values as indicated by the $\sim{}^{+60\%}_{-30\%}$ variation 
in Table \ref{tbl:scalings}. 
As our intent is to provide an estimation for advance warning times in terms of orders of magnitude, this variation is tolerable.
Moreoever, it seems that this variation tends to lengthen $\tau_\text{insp}$ more than shorten it thus our setting
$m_1=m_2=1.4 M_\odot$ is more likely to underestimate $T_\text{AW}$ than overestimate it. 
%
\item {\it Neglecting eccentricity.}
We further simplify our treatment by considering only quasi-circular inspirals meaning that at any given instant, 
the orbit can be treated as circular with the corresponding Keplerian frequency $f_K = (GM/r^3)^{1/2}/(2\pi)$. 
It was Peters and Matthews who first showed that eccentric binaries circularize in the weak field \cite{PM1963}. 
Using $\dot{a}=\dot{a}(a,e), \dot{e}=\dot{e}(a,e)$ they obtained
\be
a(e) = c_0\, \f{e^{12/19}}{1-e^2}\left(1 + \f{121}{304}e^2\right)^{870/2299} \equiv c_0\, g(e) \ , \label{eq:a_of_e}
\ee
where $a$ is the semi-major axis of the eccentric orbit and $c_0$ is a constant. For $e\ll 1$ we have $g(e)\approx e^{12/19}$, thus we may write $e\approx [a\, g(e_i)/a_i ]^{19/12}$ 
for a system with initial parameters $a_i, e_i$.
Let us apply this to the Hulse-Taylor binary pulsar \cite{Hulse:1974eb,Weisberg:2010zz} to show that the BNS orbits circularize by the time they enter the interferometers' bandwidth (ca. $1-10\,$Hz).
Letting $f =2f_K= 1\,\text{Hz}$, i.e., 
$a \simeq 3352\,$km and using the currently observed values $a_i \approx 1.95\times 10^6\,$km and $e_i \approx 0.617$,
we obtain $e\approx 6\times 10^{-5}$ at 1\,Hz. Thus our setting $e=0$ is very reasonable\footnote{
It is possible to have 
high-eccentricity binary black hole mergers in the vicinity of galactic nuclei due close encounter captures \cite{OLeary:2008myb}.
However, it is not clear how much this applies to BNS systems, nonetheless see Ref.~\cite{Chaurasia:2018zhg} for a recent numerical study of an eccentric BNS merger.}.
%
%
\item {\it Quasi-circularity.}
This implies that the timescale for the orbital radius to decrease is much longer than the orbital timescale, i.e., 
$|\dot{r}|\ll r\Omega$ during the inspiral. From $\Omega^2 = GM r^{-3}$ we can straightforwardly obtain 
%
%
an expression for the ratio of radial speed to tangential speed
\be
\f{|\dot{r}|}{r\Omega}=\f{2}{3}\f{|\dot\Omega|}{\Omega^2} \label{eq:ratio_of_radial_vel_to_tang_vel} ,
\ee
which translates the quasi-circularity condition above to $|\dot\Omega| \ll \Omega^2$ which, via Eq.~(\ref{eq:fdot}), gives
\be
\f{|\dot\Omega|}{\Omega^2} = \f{96}{5} \f{(G M_c)}{c^5}^{5/3}\Omega^{5/3} \ll 1 \label{eq:omega_dot_over_omegasq}
\ee
%
This expression yields $|\dot{\Omega}|/\Omega^2\approx 5.5\times 10^{-4}$ at $f=100\,$Hz, which, as shown above, is a few seconds
before the merger so the quasi-circularity assumption seems to
hold almost until the plunge at $f=f_\text{ISCO}$.
%
\item {\it Neglecting higher-order multipole moments.}
Another reasonable question is how much our estimation of advance warning times is
affected by neglecting higher-order contributions to the loss of energy coming from, e.g., the current quadrupole and the mass octupole moments.
The power radiated by the current
quadrupole ($\dot{E}_\text{cur}$) by a binary in a circular orbit is $\ord(v^2/c^2)$ smaller than the dominant quadrupole-mode power emission. More precisely, we have \cite{Maggiore}
\be
\f{\dot{E}_\text{cur}}{\dot{E}} = \f{1215}{896} \f{v^2}{c^2}\left(\f{m_2-m_1}{M}\right)^2=\f{1215}{896}\f{(\pi\, G M f)^{2/3}}{c^2}\left(\f{m_2-m_1}{M}\right)^2\, . \label{eq:Oct_quad_ratio}
\ee
At $f=100\,$Hz, the first two factors $\simeq 3.6\times 10^{-2}$ for $M=2.8 M_\odot$ which tells us that even in the strong-field regime, 
the next-to-leading order contribution is two orders of magnitude smaller.
The ratio in Eq.~(\ref{eq:Oct_quad_ratio}) is further supressed by $(m_2-m_1)^2/M^2$ which, for our idealized equal-mass system, 
yields zero, but if we consider $m_1=1.2 M_\odot, m_2=1.6 M_\odot$ then it gives $\approx 2\times 10^{-2}$ hence the ratio in Eq.~(\ref{eq:Oct_quad_ratio}) $\lesssim 10^{-3}$.
Thus, we can neglect the current-quadrupole radiation completely. In a similar fashion, it can be shown that the power radiated by the mass octupole ($\ell=3$)
scales the same way as in Eq.~(\ref{eq:Oct_quad_ratio}) with respect to the dominant mass quadrupole radiation, but is $\sim 50$ times smaller hence we can neglect it as well \cite{Maggiore}.
Higher moments are suppressed by higher factors of $v^2/c^2$ as can be shown using post-Newtonian theory. 
Thus, our exclusiveness to the mass-quadrupole radiation is justified.
\item {\it Neglecting neutron star spins.}
It is generally believed that neutron stars do not have large enough angular momentum to impart detectable spin effects into the GWs emitted by the BNS inspirals.
The dominant effect due to the spins is the spin-orbit (SO) precession which contributes to the energy flux at the 1.5PN order [$\ord(v^3/c^3)$] with respect to the leading-order quadrupole $\dot{E}$ of Eq.~(\ref{eq:Edot}) \cite{Blanchet_LRR}. 
Restricting to the case of $m_1=m_2$ we obtain
\be
\f{\dot{E}_{\text{SO}}}{\dot{E}} \lesssim 4 \f{G\Omega m}{c^3}\chi + \ord\left(\f{v^5}{c^5}\right) \label{eq:spin_orbit}\, ,
\ee
where we introduced the dimensionless Kerr spin parameter $\chi\equiv {S}/({G m^2})$ which, for a neutron star (NS) with spin $S$ and mass $m$, can be written as
\be
\chi=\f{c}{G} \f{I}{m^2} \f{2\pi}{P},  \label{eq:NS_spin}
\ee
where $P$ is the spin period, and $ I = (2/5) m R^2 \kappa$ is the moment of inertia of the NS with $R$ being the radius of the NS and $\kappa \sim \ord(1)$ an intrinsic constant (see, e.g., Sec.~6.2 of Ref.~\cite{Colpi_Sesana}). 
A rather compact NS with $m=1.4 M_\odot, R=11\,\text{km}, P= 10\,\text{ms}$ has $\chi\lesssim 0.05$ whereas
the fastest observed NS spin in a BNS system is $\approx 0.04-0.05$ \cite{0264-9381-26-7-073001, Stovall:2018ouw}.
Therefore, at $f= 100\,$Hz, the first term on the right-hand-side of Eq.~(\ref{eq:spin_orbit}) gives $\approx 4\times 10^{-4}$. 
On the other hand, the spin-spin coupling contributes at $\ord(v^4/c^4)\times\dot{E}$ hence $\lesssim \dot{E}_\text{SO}$.
Thus, NS spins can be neglected for our estimation of advance warning times. 
%
\item {\it Neglecting cosmological effects.}
When we take into account the effects of the large-scale structure of the universe, many quantities of interest are scaled by $(1+z)^\alpha$ where $\alpha$
is the appropriate power. The standard treatment adopts a Friedmann-Robertson-Walker cosmology with scale factor $a(t)$ and curvature $k=\pm 1, \text{ or } 0$.
Introducing the subscript $s$ to denote quantities evaluated at the source frame, we have the following cosmological corrections
\begin{align}
dt& \rightarrow (1+z) dt_s, \\
 f & \rightarrow (1+z)^{-1} f_s\, 
\end{align}
and the luminosity distance gets rescaled as
\be
D  \rightarrow  (1+z)a(t) D.
\ee
Given that $\dot{f}_s=df_s/dt_s\sim f^{11/3}_s, h_s(t)\sim D^{-1} f_s^{2/3}$, and $\tau_{\text{insp},s}\sim f_s^{-8/3}$,
the observed frequency evolution, the inspiral time, 
and the GW strain change according to 
\begin{align}
\dot{f}\ \ &\sim (1+z)^{5/3}{f}^{11/3},\\
\tau_\text{insp}& \sim (1+z)^{-5/3} \tau_{\text{insp}},\\
h(t) & \sim (1+z)^{5/3} h(t).
\end{align}
Looking at Eqs.~(\ref{eq:fdot}, \ref{eq:tau_insp}, \ref{eq:h_c}) we spot a common factor of $M_c^{\pm5/3}$ in all of them.
Therefore, we can accommodate the cosmological effects by introducing the redshifted chirp mass $\mathcal{M}_c\equiv (1+z) M_c$
which changes Eqs.~(\ref{eq:fdot}, \ref{eq:tau_insp}, \ref{eq:h_c}) to
%
\begin{align}
\dot{f} &= \f{96}{5}\pi^{8/3} \f{(G \mathcal{M}_c)}{c^5}^{5/3}\, f^{11/3} \label{eq:fdot_redshifted},\\
\tau_\text{insp}(f) &= \f{5}{256\pi}\f{c^5}{(\pi G \mathcal{M}_c)^{5/3}} \,f^{-8/3}\label{eq:tau_insp_redshifted}, \\
h_c(t) & = \f{4}{D}\left(\f{G \mathcal{M}_c}{c^2}\right)^{5/3}\left(\frac{\pi f}{c}\right)^{2/3} \label{eq:strain_TD_redshifted},
\end{align}
where $f$ is the usual GW frequency in the observer/detector frame.
So, we see that the effects of the cosmological redshifting can be accounted for by the transformation $M_c\rightarrow \mathcal{M}_c$.
This is because both $\dot{f}$ and the GW amplitude are governed by the time scale $G M_c/c^3$.
Accordingly, the inspiral evolves faster in the lab frame 
and we must now terminate it at a redshifted GW ISCO frequency which is less than $f_\text{ISCO}$ of Eq.~(\ref{eq:f_isco}).
On the other hand, the Fourier transform of $h_c(t)$ changes to 
\be
\tilde{H}(f) = A\, \f{c}{D} \f{1}{(1+z)} \left(\f{G \mathcal{M}_c}{c^3}\right)^{5/6} f^{-7/6}
\ee
indicating that the frequency-domain strain in the detector frame decreases by a factor of $(1+z)^{-1/6}$. 
For the Advanced-LIGO-Virgo network,
we consider a maximum value of 200\,Mpc for $D$,
which corresponds to $z \approx 0.02$ which redshifts $f_\text{ISCO}$ by $\approx 2\%$ and reduces $\tau_\text{insp}$ by $\approx 7\%$.
Therefore, we will neglect the effects of cosmological redshift 
when considering sources for the ALV network.
On the other hand, for Einstein Telescope, we will go out to $D=1\,\text{Gpc}\implies z\approx 0.2$
which reduces $\tau_\text{insp}$ by $\sim 25\%$, thus we shall include the redshift effects in our computations of advance warning times and SNRs
for which we provide the details in Sec.~\ref{Sec:ETB} (additional details can be found in, e.g., Ref.~\cite{Phinney:2001di}). 
\item {\it Changing the detection criterion from} $\bar{\rho}_\text{tot}=15$. 
In sections \ref{sec:results} - \ref{sec:BH_NS}, we present our results for advance warning times based on a detection criterion
given by total network SNR equalling 15 [cf. Eq.~(\ref{eq:SNR_bar})].
It is quite common in the literature to select 8 or 12 for $ \bar\rho_\text{tot} $.
So, here, we briefly explore the consequences of (i) relaxing our detection criterion to $\bar{\rho}_\text{tot}=12$, and (ii) making it more strict by setting $\bar{\rho}_\text{tot}=20$.
The former choice provides longer warning times and the latter less.
We recommend first the reading of Secs.~\ref{sec:GW170817}, \ref{sec:results} before continuing here.

For the ALV network of 2020s, it is clear from the results in Table~\ref{table:LIGO2020} that
there is no point in discussing the $\bar{\rho}_\text{tot}=20$ case as that would result in advance
warning times of mere seconds. So we only consider $\bar{\rho}_\text{tot}=12$ and BNS inspirals at $D=50,100\,$Mpc for which we find that the advance warning times increase to 120 and 36 seconds from 86 and 23, respectively. Though this is an improvement, it does not change our conclusion to Sec.~\ref{Sec:ALIGO2020}, that we expect no forecasts in the 2020s.

For Einstein Telescope, we consider both $\bar{\rho}_\text{tot}=12$ and 20 given its low-frequency sensitivity. We present the resulting advance warning times ($T_\text{AW}$) in Table~\ref{table:SNR_12_20_results}
on the next page. Comparing these with the corresponding ones in Table~\ref{table:ET}, we see that setting
$\bar{\rho}_\text{tot}=12$ increases ET-B, C warning times by $\gtrsim 50\%, \sim 20\%$, respectively.
On the other hand, $\bar{\rho}_\text{tot}=20$ decreases them by $\sim 40\%, 20\%$.
Be that as it may, we see that ET-C sensitivity still yields advance warning times $> 1\,$hour for sources within $\approx 450$\,Mpc. Therefore, our conclusions that we expect ET-C to forecast $\sim \ord(100)$ yearly
BNS mergers holds.

\begin{table}[h]
\centering
\begin{tabular}{lccccc}
\hline\hline
$D\,$(Mpc)\hspace{2mm} & \multicolumn{2}{c}{$T_\text{AW}(\bar{\rho}_\text{tot}=12)$} & & \multicolumn{2}{c}{$T_\text{AW}(\bar{\rho}_\text{tot}=20)$}\T\B\\
\hline
{}& ET-B & ET-C &\hspace{5mm} & ET-B & ET-C  \T\B\\
100 & 70\,minutes & 6.6\,hours & & 28\,minutes & 4.2\,hours \\
200 & 18\,minutes & 3.4\,hours & & 7.0\,minutes & 2.2\,hours \\
400 & 4.6\,minutes & 1.8\,hours & & 1.5\,minutes & 1.2\,hours \\
1000 & 31\,seconds & 44\,minutes & & 6.2\,seconds & 24\,minutes \\
\hline\hline
\end{tabular}
\caption{Alternative advance warning times obtained using detection criteria of $\bar{\rho}_\text{tot}=12,20$ as opposed to 15 which is our standard choice throughout this article for which the ET-B/C results are displayed in Table~\ref{table:ET}.}\label{table:SNR_12_20_results}
\end{table}
%
%
%
%
\end{enumerate}
\section{GW170817 as a test of our model for the Advanced LIGO-Virgo network}\label{sec:GW170817}
On 17 August 2017, at 12:41:04 UTC, GWs from the inspiral and merger of a binary neutron star system 
swept the ALV network's frequency band from $\approx 30\,$Hz to $\approx 2000\,$Hz, accumulating a total SNR of
$\rho_\text{tot} = 32.4$ \cite{GW170817}.
The inferred duration of the event in the network bandwidth was $\approx 57\,$ seconds during which $\approx 3000$ cycles of GWs were emitted \cite{GW170817_Facts}.
Assuming low-spin priors  ($\chi \le 0.05$, see point 7 of Sec.~\ref{sec:idealizations}), the following updated values were inferred for the total mass, the chirp mass,
and the luminosity distance of the source with 90\% confidence \cite{GW170817_2018}:
\be
M = 2.73^{+0.04}_{-0.01} M_\odot,\quad M_c = 1.186^{+0.001}_{-0.001} M_\odot,\quad D= 40.7^{+2.36}_{-2.36}\,\text{Mpc}. \label{eq:GW170817_params}
\ee
Note that the error bar for the chirp mass is very small
due to the fact that it can be extracted to high precision over thousands of GW cycles via Eq.~(\ref{eq:fdot}). 
%
%

How well does the ``naive'' Newtonian evolution based on the quadrupole formula perform? 
To test this, let us first compute the inspiral time, $\tau_\text{insp}$, and the number of GW cycles, $\mathcal{N}_\text{cyc}$, for this BNS system with initial frequency $f_0=30\,$Hz and final frequency $f_\text{ISCO} \simeq 1605\,$Hz. 
Eqs.~(\ref{eq:tau_insp}) and (\ref{eq:Ncyc1}) provide us with the leading-order values
\be
\tau_\text{insp}^{0\text{PN}} \simeq 56.06\,\text{seconds}, \quad \mathcal{N}_\text{cyc}^{0\text{PN}} \simeq 2687\,\label{eq:GW170817_values_0PN}. 
\ee
If we wish to do better, we can add all the post-Newtonian contributions up to 2PN resulting in
\be
\tau_\text{insp}^{2\text{PN}} \simeq 57.06\,\text{seconds}, \quad \mathcal{N}_\text{cyc}^{2\text{PN}} \simeq 2735 \label{eq:GW170817_values_3p5PN}\, ,
\ee
which more or less match the detected inspiral time and number of cycles. We stop at 2PN because the currently available 3.5PN expressions require the value of $\nu$
which is not well known for GW170817.
%

We now put our model three-IFO network to the test
by computing the SNRs accumulated by each individual IFO and by the entire network. To do so, we take the originally-inferred values for $M,M_c,D$ from GW170817 given in Eq.~(\ref{eq:GW170817_params}) above and compute the SNRs $\rho_i, i=h,m,l$ via a slightly modified version of Eq.~(\ref{eq:SNR_i})
\begin{align}
\bar{\rho}_i\equiv\rho_i(f_0,f_f) &= 2\tilde{A}_i\, h_0\left[ \int_{f_0}^{f_f} df\, \f{f^{-7/3}}{S_\text{n}(f)}\right]^{1/2} \label{eq:SNR_i_with_fi}.
\end{align}
where $f_0 =30\,\text{Hz}, f_f= f_\text{ISCO}$ for GW170817 and $\tilde{A}_i$ given in Table~\ref{table:network_params}.
For this computation, we have neglected the cosmological corrections as 40\,Mpc is in our ``local'' neighbourhood.

For $S_\text{n}(f)$ we use actual [noise-subtracted] LIGO-Livingstone (L1) data for GW170817 from LIGO's website \cite{LIGO_L1_data}.
The data, sampled at 4096\,Hz, is given as a discrete time series $h(t_i)$ for a duration of 2048\,seconds in time steps of $\Delta t=1/4096\simeq 2.44\times 10^{-4}\,$seconds.
We first Fourier-transform the data to frequency space and apply several high and low-frequency filters to isolate a window of $f\in [10, f_\text{ISCO}]$.
The LIGO Open Science Center has a detailed Python-based tutorial on how to do this using the data from the very first GW detection (GW150914) \cite{LIGO_tutorial}. 
We use a $\verb|Mathematica|$ based code developed at University College Dublin (UCD) to do the same for GW170817.
\begin{figure}[ht!]
\includegraphics[height=9cm]{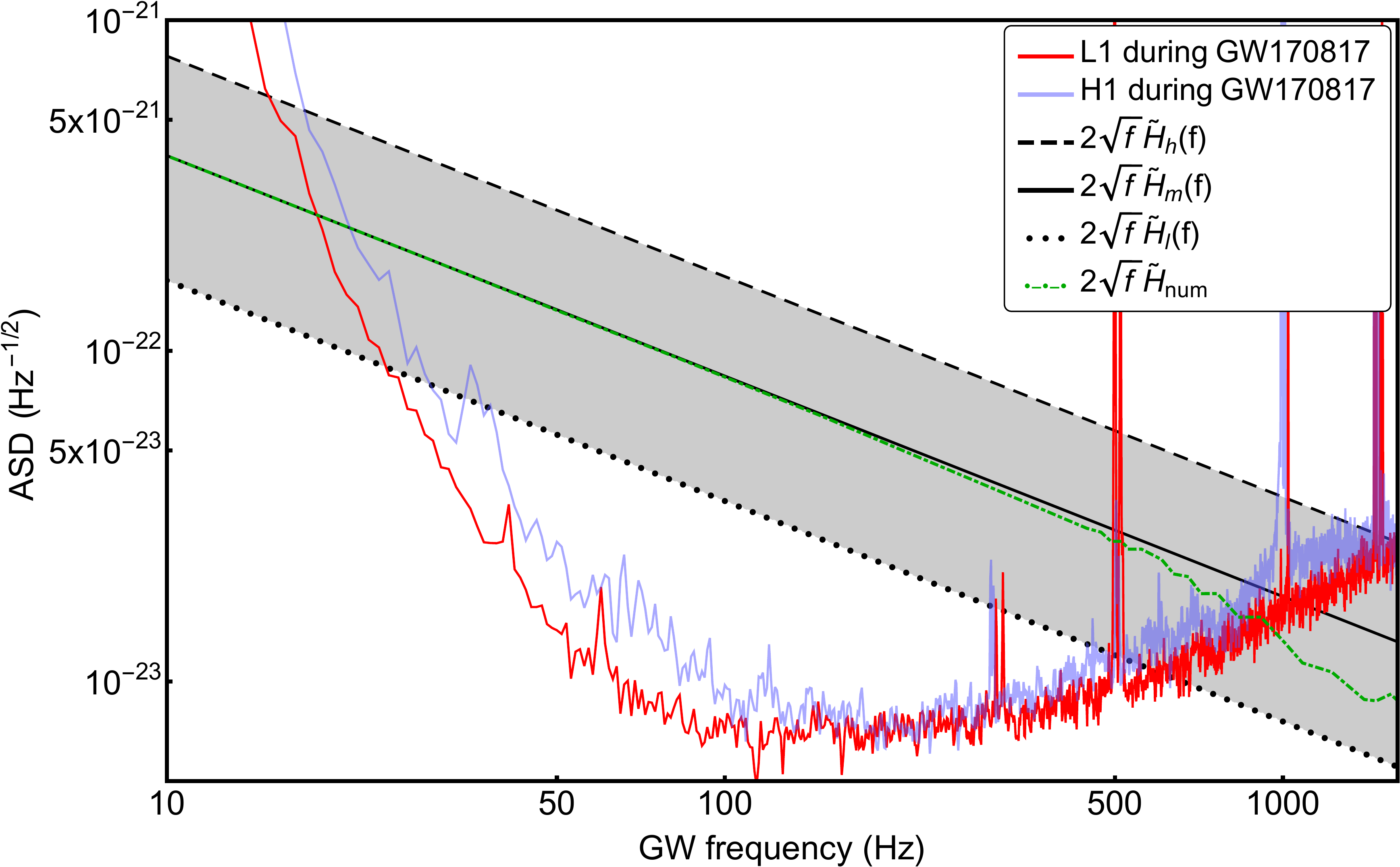}
\caption{GW170817 as it sweeps from 10\,Hz to $f_\text{ISCO} \simeq1605\,$Hz across our network with ASD (sensitivity) of 
LIGO Livingstone (L1) during the event (the darker red curve).
We also show the Hanford detector's sensitivity in light blue for comparison. The dashed, solid and dotted lines represent the 
frequency-domain GW strains [$2\sqrt{f} \tilde{H}_i(f)$ with $i=h,m,l$] due to the inspiral of
a binary neutron star system with parameters given in Eq.~(\ref{eq:GW170817_params}).
The variation in strain amplitudes is the result of a given interferometer's orientation and the sky position of the source 
as explained in Sec.~\ref{sec:LIGO_topo} and summarized in Table~\ref{table:network_params}.
More specifically, the dashed line is how ``loud'' GW170817 would appear in the data stream of an IFO with L1's sensitivity
with near optimal orientation
(quality factor $|Q|=4/5=80\%$). Similarly, the solid line represents the strain in an identical interferometer with 
$|Q|=2/5=40\%$ (RMS-averaged response). 
Finally, the dotted line represents a pessimistic IFO orientation and source location (sub-optimal configuration)
resulting in a reduction of $|Q|$ to $\simeq 16.7\%$.
The shading represents the region of likelihood for the strain due to an inspiralling BNS with parameters of Eq.~(\ref{eq:GW170817_params}). In the case of a three-IFO
network, we would expect all three strains to exist somewhere in the shaded region most of the time for an inspiral with GW170817's intrinsic parameters. 
The dot-dashed (green) curve shows a more realistic strain constructed with data from numerical simulations of Ref.~\cite{Read:2009yp}
and rescaled to match our solid line at $f=10\,$Hz.
The breakdown of the $\sqrt{f}f^{-7/6}=f^{-2/3}$ behavior of
the strain around $f\sim100\,$Hz due to tidal effects is clear.
}
\label{fig:figGW170817}
\end{figure}\\
Fig.~\ref{fig:figGW170817} shows our network. The red curve is L1's ASD (sensitivity) during GW170817. We assume our three IFOs have identical sensitivities,
but because their orientations and response functions differ, their sensitivity to a localized source is reduced. We represent this reduction by lowering the
amplitude of the GW strain by appropriate quality factors ($|Q_i|$) explained in Sec.~\ref{sec:BNS_inspiral} and shown in Table~\ref{table:network_params}.
The amplitude-adjusted strains $ \tilde{H}_i(f)$ are represented by the dashed ($i=h$), solid ($i=m$), and dotted ($i=l$) lines in the figure where we actually plot
$2\sqrt{f} \tilde{H}_i(f)$ as is standard. 
Using Eq.~(\ref{eq:SNR_i_with_fi}) we can immediately compute the SNRs accumulated by our network
\be
\bar\rho_h = 28.7,\quad \bar\rho_m = 14.3,\quad \bar\rho_l = 6.00 \label{eq:GW170817_SNRs}.
\ee
These are comparable to actual SNRs of L1, H1, and V1 for GW170817, which are, 26.4, 18.8, and 2.0, respectively\footnote{At the time of GW170817, Virgo's sensitivity had not reached the level of LIGO's. Hence it accumulated a much smaller SNR for GW170817. By 2020, Virgo should be operating at nearly the same sensitivity of LIGO.}.
The shaded region in Fig.~\ref{fig:figGW170817} covers strains between $\tilde{H}_l(f)$ and $\tilde{H}_h(f)$. 
Given a randomly positioned/oriented BNS inspiral at 40 Mpc, we would in general expect a randomly oriented IFO's response to lie in this shaded region.
%

%
%

We can now compute the total SNR of GW170817 as inferred by our network
\be
\bar\rho_\text{tot}= \sqrt{\bar\rho^2_h+\bar\rho^2_m+\bar\rho^2_l}= 32.0 \, \label{eq:SNR_total_GW170817} .
\ee
This matches the actual value of 32.4 very well. Therefore, we conclude that our model network performs well enough to 
act as our future ALV network of the 2020s once we change
the ASD from the August 2017 value to design sensitivity.
%
%
%
%
%
\section{Results: Advance warning times for future binary neutron star inspirals}\label{sec:results}
 Our aim is to forecast GRBs using our model networks for the ALV era in the 2020s
 and the Einstein Telescope era in the 2030s.
 Our EM follow-up capabilities will depend on how much advance warning the future networks will give us prior to the merger/GRB.
 Our definition of advance warning is the inspiral time to the merger from a certain threshold instant $\bar{t}$ at which point the network ``agrees'' that there is a GW transient
 statistically significant enough to issue warnings to electromagnetic telescopes.
 We define this threshold instant to be given at a frequency $\bar{f}$ when the total SNR equals 15. 
 More specifically, in the case of the ALV network, our statement is that there exists a frequency $\bar{f} < f_\text{ISCO}$ such that
 \be
 \bar\rho_\text{tot}= \sqrt{\left[\rho_h(f_0^h,\bar{f})\right]^2+\left[\rho_m(f_0^m,\bar{f})\right]^2+\left[\rho_l(f_0^l,\bar{f})\right]^2}=15, \label{eq:SNR_bar}
 \ee
 where $f_0^i$ are the initial GW frequencies at which a given inspiral enters the $i^\text{th}$ IFO's sensitivity band, and
 are given by the solution set to
 \be
 \sqrt{S_\text{n}(f_0^i)} = 2\sqrt{f_0^i}\, \tilde{H}_i(f_0^i) \label{eq:f0},\qquad i=h,m,l
 \ee
 with $f_0^i < \bar{f}$.
 Once we know $\bar{f}$, we can compute the remaining time to the merger, $\tau_\text{insp}(\bar{f})$, using Eq.~(\ref{eq:tau_insp})
 and its 3.5PN-enhanced version from Ref.~\cite{Blanchet_LRR}. This is what we call our advance warning time $T_\text{AW}$.
 For simplicity, we set $m_1=m_2=1.4 M_\odot \implies M_c\simeq 1.219 M_\odot$, which we justified in
 Point 3 of Sec.~\ref{sec:idealizations}. 
 With the masses fixed, the only remaining variable is the 
 luminosity distance $D$.
For the ALV network, we consider $D=50,100,200\,$Mpc 
 as A-LIGO's range will be slightly over 200\,Mpc and A-Virgo's around 130\,Mpc.
 For each value of $D$, we compute $\bar{f}$ from Eq.~(\ref{eq:SNR_bar}) with which we then compute $T_\text{AW}=\tau_\text{insp}(\bar{f})$.
 We summarize our procedure in the flow diagram below.
  \begin{figure}[ht!]
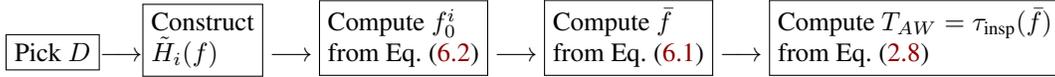

 \framebox{\Longstack[l]{Pick $D$}}$\longrightarrow$\framebox{\Longstack[l]{Construct\\ $\tilde{H}_i(f)$}} 
 $\longrightarrow$  \framebox{\Longstack[l]{Compute $f_0^i$\\ from Eq.~(\ref{eq:f0})}} $\longrightarrow$  \framebox{\Longstack[l]{Compute $\bar{f}$\\ from Eq.~(\ref{eq:SNR_bar})}}
 $\longrightarrow$  \framebox{\Longstack[l]{Compute $T_{AW}=\tau_\text{insp}(\bar{f})$\\ from Eq.~(\ref{eq:tau_insp})}}
  \caption{Flow diagram summarizing our procedure for computing advance warning times $T_\text{AW}$ with the Advanced-LIGO-Virgo network. 
  We compute $T_\text{AW}$ for BNS systems with $m_1=m_2=1.4M_\odot$ inspiralling at $D=50,100,200\,$Mpc. 
  }\label{fig:flow}
\end{figure}

\begin{figure}[ht!]
\includegraphics[width=\linewidth]{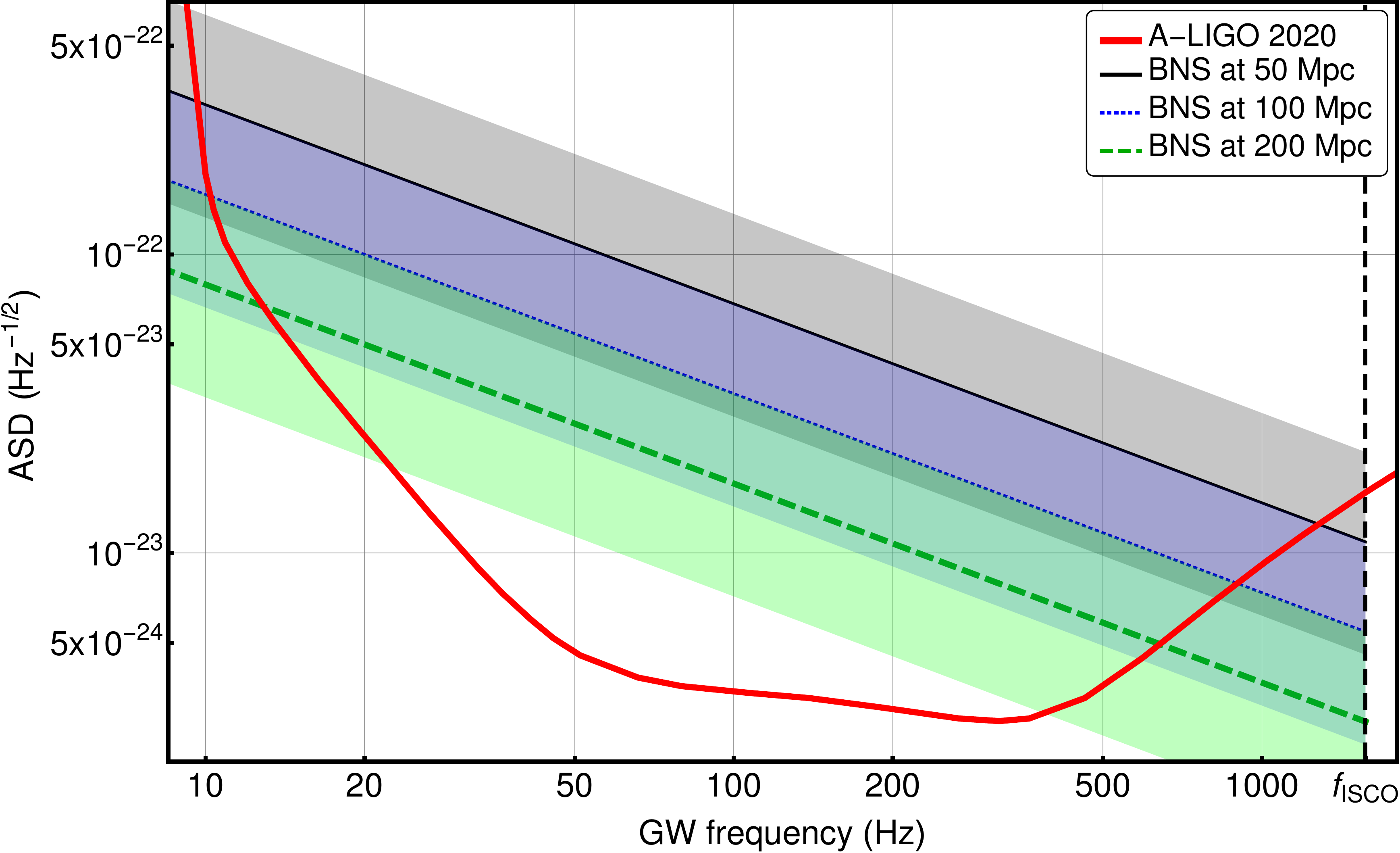}
\caption{Typical $m_1=m_2=1.4 M_\odot$ BNS inspirals sweeping across the Advanced LIGO-Virgo network in the 2020s
which may be the harbingers of short-hard GRBs.
The solid (black), dotted (blue), and dashed (green) lines are the RMS-averaged strains ($2\sqrt{f}\tilde{H}_m$) at luminosity distances of $D=50, 100, 200\,$Mpc, respectively. The accompanying shaded regions span from the near-optimally oriented IFO 
($2\sqrt{f}\tilde{H}_h$) to
sub-optimally oriented IFO ($2\sqrt{f}\tilde{H}_l$) as explained in Fig.~\ref{fig:figGW170817}.
The vertical dashed (black) line marks the ISCO frequency $f_\text{ISCO} \simeq 1571\,$Hz at which point we terminate the inspirals. The light red curve shows the L1 ASD during GW170817.
}
\label{fig:LIGO2020}
\end{figure}

%
 %
 %
  %
%
We additionally calculate 
the total accumulated SNR from $f_0$ to $f_\text{ISCO}$ for each inspiral using
\be
\bar{\rho}_\text{F}\equiv\sqrt{\left[{\rho}_h(f_0^h,{f}_\text{ISCO})\right]^2+\left[{\rho}_m(f_0^m,{f}_\text{ISCO})\right]^2+\left[{\rho}_l(f_0^l,{f}_\text{ISCO})\right]^2} \label{eq:SNR_bar_FINAL}\, ,
\ee
where $f_\text{ISCO}$ is given by Eq.~(\ref{eq:f_isco}).
We will introduce the corresponding expressions for ET in Sec.~\ref{Sec:ETB}.
To compute the SNRs defined above for the ALV network we require IFO noise data for which
we use the ASD for the BNS-optimized A-LIGO design sensitivity expected to be attained ca.~2020 from Ref.~\cite{LIGO2020},
which is plotted as the thick solid (red) curve in Fig.~\ref{fig:LIGO2020}.
Sweeping across the detector's frequency band are RMS-averaged GW strains ($\tilde{H}_m$) due to BNS inspirals at $D=50,100,200\,$Mpc plotted from top to bottom, respectively (recall we plot $2\sqrt{f}\tilde{H}_i$ against detector noise).
Each RMS-averaged strain is accompanied by its shaded region ranging from a near-optimally oriented strain ($\tilde{H}_h$) to a sub-optimally oriented strain ($\tilde{H}_l$) which were explained in Sec.~\ref{sec:GW170817} and first shown in Fig.~\ref{fig:figGW170817}.
The height of a given strain above the detector sensitivity provides a good \emph{visual} estimation for the corresponding SNR,
but the reader should keep in mind that the actual computation involves the integral of a ratio, not a difference [cf. Eq.~(\ref{eq:SNR})].
\subsubsection{Forecasting GRBs in the 2020s with the Advanced LIGO-Virgo Network}\label{Sec:ALIGO2020}
%
%

The advance warning time computation follows the procedure outlined in Fig.~\ref{fig:flow}. We use numerical root finding to determine $\bar{f}$ and $f_0^i$ for all three values of $D$. 
The computation of SNRs and $T_\text{AW}$ is then straightforward. 
We summarize our results in Table \ref{table:LIGO2020} where we present both the 0PN and the 3.5PN results for the advance warning times
$T_\text{AW}$. As can be seen from the values for $T_\text{AW}$ in the table, we can at best expect 85\,seconds of early warning time.
This time increases to approximately two\,minutes in the case of another GW170817-like event with $D=40\,$Mpc.
We can estimate the likelihood of such an event happening again
using the current best estimate for the BNS inspiral rate 
inferred from the O1 and O2 observing periods of Advanced LIGO: $ R=1540^{+3200}_{-1220}\,$Gpc$^{-3}$yr$^{-1}$ \cite{GW170817}.
This translates to $\approx 0.1$ event per (40-Mpc)$^3$ per year with the upper limit $\approx 0.3$. So we can at best hope to detect
one 40-Mpc inspiral per a three-year period.
Given that we might at best have a couple of minutes of warning time for a potentially once-a-decade event, we do not expect to be able forecast GRBs in the Advanced LIGO-Virgo era. 
%
%
\begin{table}[h]
\centering
\begin{tabular}{lcccccc}
\hline\hline
$D\,$(Mpc) & $\bar{f}\,$(Hz) &{}& $T^{0PN}_\text{AW}$(sec) &\hspace{1mm} & $T^{3.5PN}_\text{AW}\text{ (sec)}\hspace{2mm}$& $\bar{\rho}_F$\T\B\\
\hline
50 & $\approx\,$25.3 & & 84.25 & & 85.86 & 70.3 \T\\
100 & $\approx\,$41.6 & & 22.39 & & 22.86 & 35.2 \\
200 & $\approx\,$121 & & 1.291 &\quad & 1.300 & 17.6 \\
\hline\hline
\end{tabular}
\caption{The ALV network's potential for providing advance warning times, $T_\text{AW}$, 
to electromagnetic observatories for the case of $m_1=m_2=1.4 M_\odot$ BNS inspirals at luminosity distances of $D=50,100,200\,$Mpc. 
$\bar{f}$ is the frequency at which the  
SNR accumulated by the network reaches 15. 
$\bar{\rho}_F$ is the total accumulated network SNR. 
We present both the Newtonian (0PN) 
and the 3.5PN result for the advance warning times.}\label{table:LIGO2020}
\end{table}
Be that as it may,
the ALV network will accumulate impressive SNRs 
(see $\bar{\rho}_F$ in Table~\ref{table:LIGO2020}) for BNS inspirals out to 100\,Mpc.
As the measurement errors for the system parameters are proportional to $\bar\rho^{-1}_F$ \cite{Cutler:1994ys},
inspirals that yield high SNRs ($\bar\rho_F\gtrsim 40$) will enable (i) high precision measurements of 
$M_c, M$ \cite{Farr:2015lna}; (ii) source sky localization to $\lesssim 5\,\text{deg}^2$ \cite{Rodriguez:2013oaa};
(iii) $ \lesssim 10\%$ precision in the inferred NS masses \cite{Rodriguez:2013oaa} and the effective spin parameter $\chi_\text{eff}$ \cite{Zhu:2017znf}; and
(iv) restrictions on the equation of state and the tidal deformability
parameters of neutron stars \cite{Read:2009yp, Andersson:2009yt, PhysRevD.89.103012, PhysRevD.91.043002}.
We should caution the reader that such high-SNR systems will make up $\lesssim 1\%$ of the population of BNSs detected by 
the ALV network \cite{Sathyaprakash:2012jk}.

In mid 2020s LIGO will first be upgraded to A+ then to Voyager. 
However, both versions will have comparable sensitivities to Advanced LIGO
at low frequencies (see Fig.~1 of Ref.~\cite{Mills:2017urp}).
Therefore, they will not start accumulating significant SNR until $f\approx 10\,$Hz, thus offering us gains of a few minutes over $T_\text{AW}$ in Table~\ref{table:LIGO2020}. For this reason, we move on to 2030s: the 
era of Einstein Telescope.

\subsubsection{Forecasting GRBs in the 2030s using the Einstein Telescope}\label{Sec:ETB}
We now wish to quantify the forecasting capabilities of Einstein Telescope. To this end, we consider the inspiral of BNS systems
at luminosity distances of $100,200,400,1000\,$Mpc entering the ET band. 
These correspond to cosmological redshifts of $z\approx 0.0222,0.0437, 0.085, 0.198$, respectively.
To compute these, we used a flat $\Lambda$CDM model ($\Omega_k=0$) with the latest Planck satellite parameters: 
$\Omega_\Lambda = 0.6911, \Omega_m = 0.3089, H_0 = 67.74\,$km\,s$^{-1}\,$Mpc$^{-1}$ \cite{Planck2015}. 
It is then straightforward to translate $D$ to $z$ (cf. Ref.~\cite{Hogg:1999ad}).

\begin{figure}[ht!]
\includegraphics[width=\linewidth]{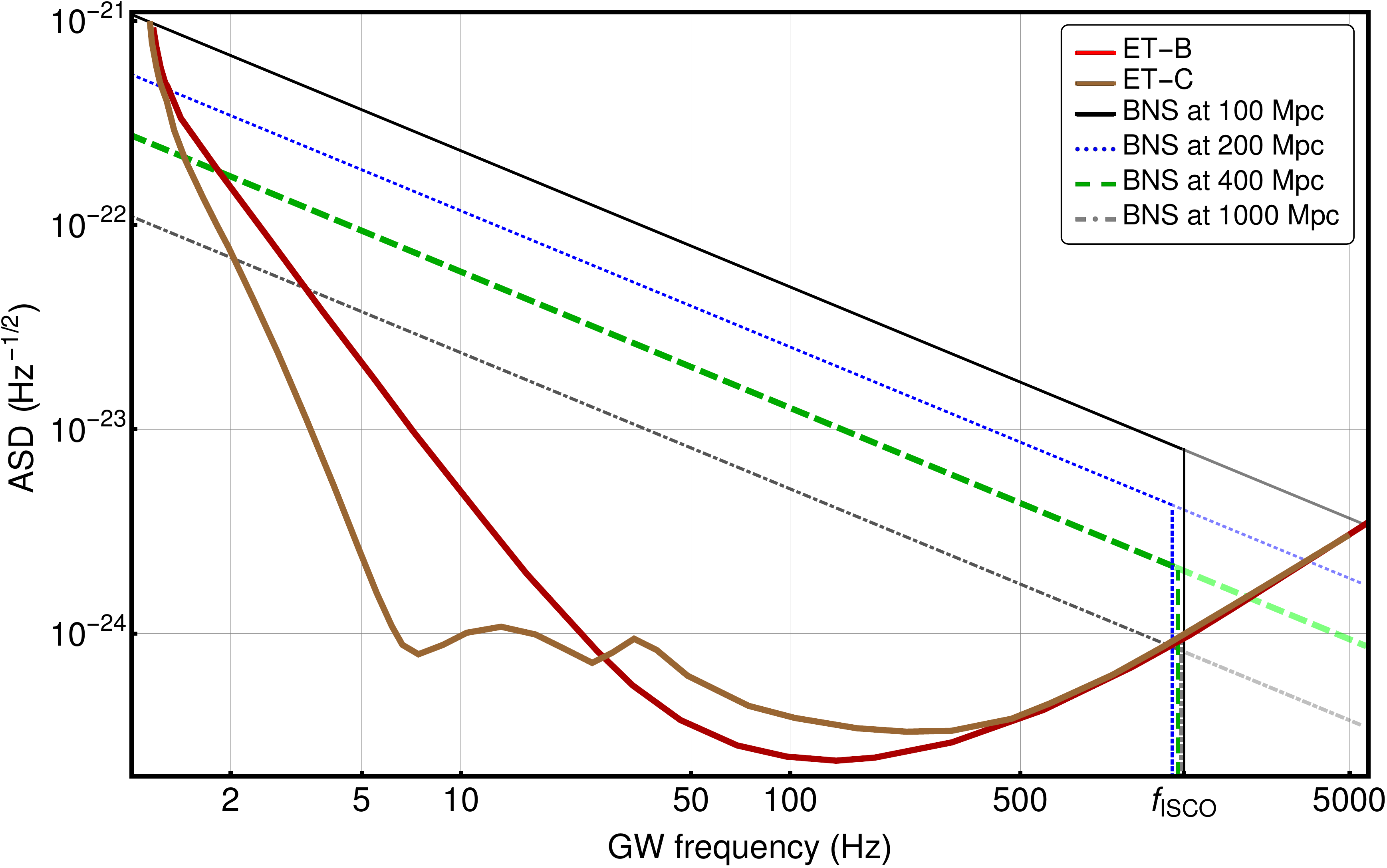}
\caption{Typical GW sources that may be harbingers of GRBs in the 2030s: $1.4 M_\odot-1.4 M_\odot$ inspiralling BNS systems sweeping across 
the Einstein Telescope's sensitivity band for both B and C configurations.
The solid (black), dotted (blue), dashed (green), and dot-dashed (gray) lines are the redshift-corrected
RMS-averaged strains, $2\sqrt{f}\tilde{H}_\text{ET}$, at luminosity distances of $D=100, 200, 400, 1000\,$Mpc, respectively. 
The vertical lines with correspondingly identical patterns (colors) mark the redshifted ISCO frequencies $(1+z)^{-1} f_\text{ISCO}$ at which point we terminate each inspiral.
As the true ISCO frequency is likely larger than $f_\text{ISCO}$ \cite{Marronetti:2003hx}, the inspirals would continue to nearly 2\,kHz indicated by the 
faded lines in the plot (drawn to ca. 5\,kHz for aesthetic reasons).
}
\label{fig:ETB2030}
\end{figure}

The SNR accumulated in ET from each inspiral is obtained by a modification to Eq.~(\ref{eq:ET_SNR})

\be
\bar\rho_{\text{ET}}(f_0,f)\equiv \f{6}{5}A h_0\, (1+z)^{-1/6}\left[\int_{f_0}^{f} d f'\, \f{f'^{-7/3}}{S^\text{ET}_n(f')}\right]^{1/2} \label{eq:ET_SNRv2},
\ee
where $f_0$ is the GW frequency at which the emitted GWs enter ET's detection band, i.e, $f_0 < f_\text{ISCO}$ such that
 \be
 \sqrt{S^\text{ET}_\text{n}(f_0)} = 2\sqrt{f_0}\, \tilde{H}_\text{ET}(f_0) \label{eq:f0_ET}\,
 \ee
with $\tilde{H}_\text{ET}(f)$ given by the redshifted version of Eq.~(\ref{eq:H_delta})
\be
\tilde{H}_\text{ET}(f) = \f{1}{2}\sqrt{\f{3}{10}}\,\pi^{-2/3} \f{c}{D}\f{1}{(1+z)}\left(\f{G \mathcal{M}_c}{c^3}\right)^{5/6} f^{-7/6}\, , \label{eq:H_delta_redshifted}
\ee
where recall $\mathcal{M}_c = (1+z)M_c$.
For the detector noise $\sqrt{S^\text{ET}_n(f)}$, we adopt the ASD for the B and C configurations known as ET-B and ET-C, respectively \cite{Hild:2010id}, which we plot as the solid dark (red) and light (brown) curves in Fig.~\ref{fig:ETB2030}.
Sweeping across these are four BNS inspirals at $D=100,200,400,1000\,$Mpc represented by solid (black), dotted (blue), dashed (green),
and dot-dashed lines (gray), respectively. These strains sit much higher than the detector noise
compared with those in Fig.~\ref{fig:LIGO2020} which foretells us that the SNRs accumulated in the ET band will be much higher, 
thus yielding much longer advance warning times as we show in Table~\ref{table:ET}.

To determine $T_\text{AW}$ we once again impose the threshold SNR of 15 which, the reader may recall, 
is our detection criterion for sending out warnings to electromagnetic observatories. Thus, rewriting Eq.~(\ref{eq:SNR_bar}) in the case of ET, we define $\bar{f}_\text{ET}$ as the solution to
\be
\bar\rho_\text{ET}(f_0,\bar{f}_\text{ET}) = 15 \label{eq:ET_fbar},
\ee
which then gives us our redshift-incorporated advance warning time via $T_\text{AW}=\tau_\text{insp}(\bar{f}_\text{ET})$ using Eq.~(\ref{eq:tau_insp_redshifted}) which we can slightly improve by using the 3.5PN expression. 
Therefore, we proceed as summarized in Fig.~\ref{fig:flow} and compute the 0PN and 3.5PN advance warning times 
for all four inspirals in both the ET-B and ET-C bands.
The 3.5PN computation requires rescaling of the PN parameters $x$ and $\Theta$ to $(1+z)^{2/3} x$ and $(1+z)^{-1}\Theta $ (see Sec.~9.3 of 
Blanchet's review \cite{Blanchet_LRR} for details).
The final accumulated SNRs are given by
\be
\bar\rho_{F}\equiv \bar\rho_\text{ET}(f_0,f_\text{ISCO})  .\label{eq:rhoF_ET}
\ee
where $f_\text{ISCO}$ is now the redshifted version of Eq.~(\ref{eq:f_isco}), equalling $1537, 1505, 1448, 1311\,$Hz, respectively for $D=100,200,400,$ 1000\,Mpc.
Table~\ref{table:ET} summarizes our results for both ET-B and ET-C where we chose to only display the 3.5PN-accurate inspiral time as $T_\text{AW}$,
but we computed this quantity also using both 0PN and 2PN expressions as a consistency check.

\begin{table}[h]
\centering
\begin{tabular}{lccccccc}
\hline\hline
$D\,$(Mpc) & \multicolumn{3}{c}{ET-B} &  & \multicolumn{3}{c}{ET-C}\T\B\\
\hline
{}& $\bar{f}_\text{ET}\,$(Hz) & \ \hspace{7mm} $T_\text{AW}$ \ \hspace{7mm} & $\bar{\rho}_{F}$ &{} & $\bar{f}_\text{ET}\,$(Hz) & \ \hspace{5mm} $T_\text{AW}\hspace{12mm}$& $\bar{\rho}_{F}$\T\B\\
100 & $\approx\,$6.72 &  47.0\,minutes & 306 &{\qquad} & $\approx\,$3.27 & 5.34\,hours\ & 365\T\\
200 & $\approx\,$11.2 & 11.6\,minutes & 152 &{\qquad} & $\approx\,$4.10 & 2.87\,hours\ & 182 \\
400 & $\approx\,$18.2 & 3.00\,minutes & 75.7 &{\qquad} & $\approx\,$5.06 & 1.51\,hours\ & 90.5\\
1000 & $\approx\,$41.3 &17.2\,seconds & 29.8& \qquad &    $\approx\,$6.76 & 35.6\,minutes & 35.6  \\
\hline\hline
\end{tabular}
\caption{Forecasting capabilities of Einstein Telescope. 
ET-B and ET-C refer to the different configurations shown in Fig.~\ref{fig:ETB2030}. For the advance warning times, we only present the result of the more accurate 3.5PN computation. $\bar{f}_\text{ET}$ is the threshold frequency at which
ET-B/C accumulate SNR of 15 which we take to be our detection criterion. Note that both $T_\text{AW}$ and $\bar\rho_F$ are larger for ET-C
due to its improved sensitivity in the $1\,\text{Hz}\lesssim f\lesssim 30\,$Hz regime compared to ET-B as is clear in Fig.~\ref{fig:ETB2030}.
These results and those of Table.~\ref{table:LIGO2020} are summarized in Fig.~\ref{fig:summary}.}\label{table:ET}
\end{table}
As the results of Table~\ref{table:ET} indicate, ET-B will provide $\gtrsim 10\,$ minutes of warning time for BNS inspirals at $D\lesssim 200\,$Mpc.
Instead of speculating whether or not this is enough to forecast GRBs, we turn our attention to ET-C's forecasting capabilities.
As can be seen from Fig.~\ref{fig:ETB2030}, ET-C's increased sensitivity in the $1\,\text{Hz}\lesssim f\lesssim 30\,$Hz regime will significantly lengthen the advance warning times.
This is best highlighted by the sixth colum of Table~\ref{table:ET}, 
where we show that ET-C will provide up to \emph{five hours} of early warning time for nearby ($D\lesssim 100\,$Mpc) neutron star mergers.
Moreover, BNS inspirals as far as 200\,Mpc (the ALV network's range) will persist in the ET-C sensitivity band for a few hours after their initial detection ($T_\text{AW} > 2\,$ hours).
Current BNS merger rate, inferred from A-LIGO's O1-O2 periods, implies that $\sim 40$ to $\sim600$ sources within a volume of $(200\,\text{Mpc})^3$ will be detected yearly by ET-C. 
Our values for $T_\text{AW}$ for $D=200, 400\,$Mpc are roughly consistent
with the findings of Ref.~\cite{Chan:2018csa}.

Given Moore's law and new search approaches (e.g., see Ref.~\cite{Gabbard:2017lja} for one based on deep learning) 
we speculate that the advances in the detection, localization and early-warning algorithms in the next decade will reduce the required advance
warning time to less than an hour; thus we set $T_\text{AW} = 1\,\text{hour}$ and obtain corresponding BNS ranges of $D_\text{BNS}=87$ and $613\,$Mpc ($z\approx 0.02,0.127$) for ET-B and ET-C, respectively.
In other words, all inspiralling BNSs closer than $D_\text{BNS}$ will give us more than an hour's leeway before the GRB (employing the detection criterion $\text{SNR} =15$ as before).
The volume set by $D_\text{BNS}$ translates to a BNS merger rate of $\sim\mathcal{O}(1)\,\text{yr}^{-1}$ for ET-B and
$\approx 355^{+730}_{-280}\,\text{yr}^{-1}$ for ET-C. Additionally, each source within $D_\text{BNS}$ will accumulate $\bar\rho_F \gtrsim 420, 58$ over its inspiral
in the ET-B, ET-C bands, respectively. These results are summarized in Table~\ref{table:horizon}.

\begin{figure}[ht!]
\includegraphics[width=\linewidth]{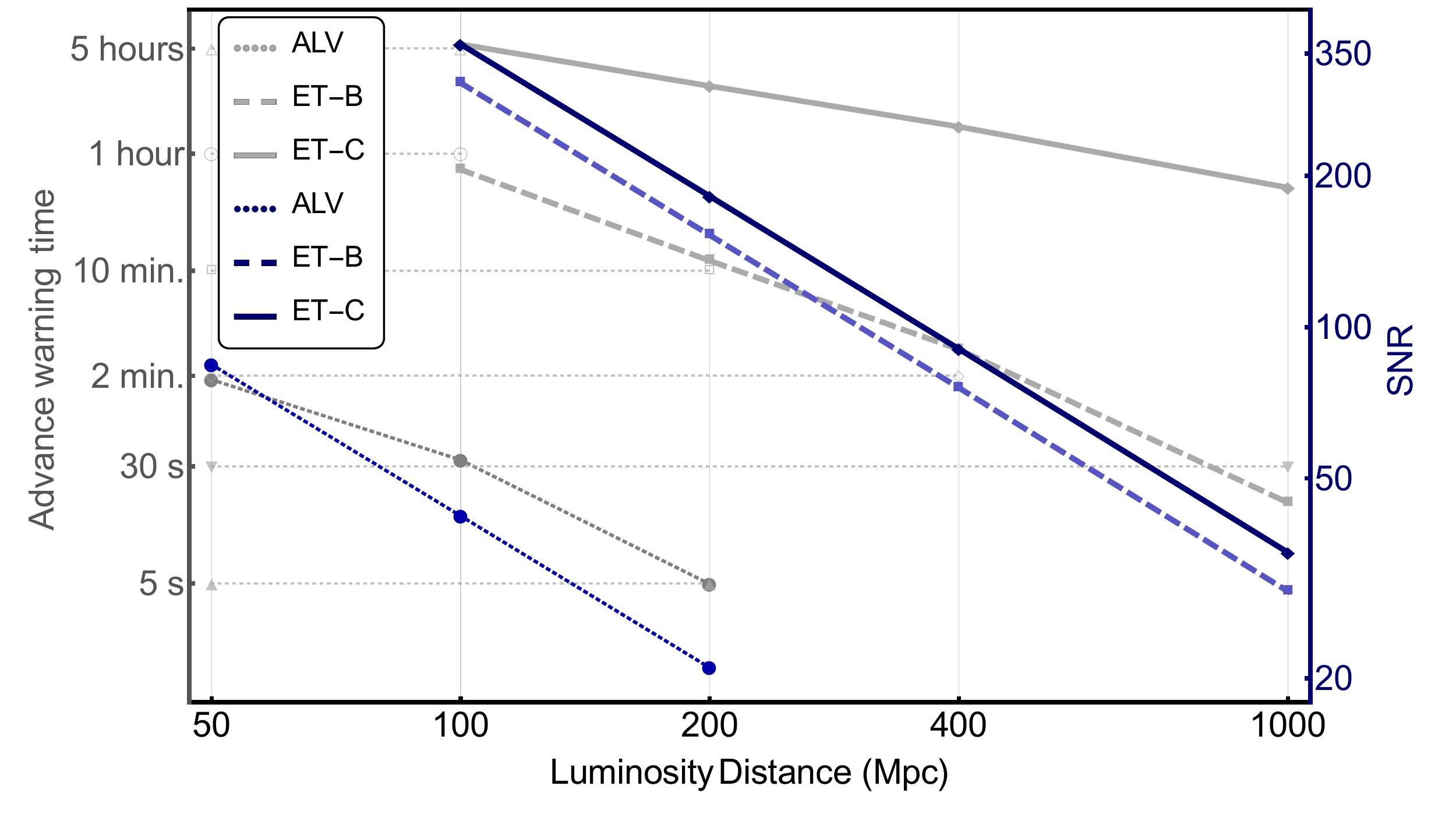}
\caption{Summary of our major findings for both the Advanced-LIGO-Virgo (ALV) network circa 2020 and the Einstein Telescope's B and C configurations
ca. 2030. The light coloured (gray) data shows $T_\text{AW}$ for the ALV2020 (dotted), ET-B (dashed), and ET-C (solid), respectively, with the corresponding axis on the left.
The connecting light (gray) lines are drawn only for visualization purposes unlike the dark (blue) lines coming from the $D^{-1}$ powerlaw of the SNR [see Eq.~(\ref{eq:ET_SNRv2})] which is plotted on the right axis. 
We used $D=50,100,200\,$Mpc to obtain the ALV results (Sec.~\ref{Sec:ALIGO2020}) and $D=100,200,400,1000\,$Mpc for the ET results.}\label{fig:summary}
\end{figure}

Both Tables~\ref{table:ET} and \ref{table:horizon} clearly show that ET will yield superb SNRs for sources out to 1\,Gpc. 
With such high SNRs we expect that a future network consisting of ET \emph{and} LIGO Voyager will be able to localize half the sources 
inside a radius of $D\approx 1000\,$Mpc to within $ 10\,\text{deg}^2$ \cite{Mills:2017urp}, but mostly \emph{after} the merger.
On the other hand, ET together with CE will localize 
$\sim 80, 20\%$ of the BNSs
within 200, 400\,Mpc to $1\,$deg$^2$, respectively \cite{Chan:2018csa}.
This is thanks to CE's low-frequency sensitivity which is sufficient to start accumulating SNR at $f\approx 5\,$Hz from BNSs within 400\,Mpc corresponding to $T_\text{AW}\gtrsim 1.5\,$hours.
Even without the help of a second detector, ET-C will be able
to localize $\sim 5\%$ of these transients to within $10\,$deg$^2$ using
the Earth's rotation because these nearby sources will spend hours in its frequency band \cite{Zhao:2017cbb}.
This percentage amounts to $\sim 5$ events per year.
Therefore, erring on the side of optimism, we conclude that the 2030s\footnote{As the ET construction/commissioning timeline is currently uncertain, we can only speculate that ET-C will be operational in mid 2030s,
which is roughly in line with the timeline of the milli-Hertz space interferometer LISA \cite{Audley:2017drz} and CE \cite{LIGO_inst}.} 
hold in store 
(i) forecasts of $\gtrsim \ord(100)$ short gamma-ray bursts and
(ii) real-time electromagnetic observations of $\gtrsim \ord(10)$ of these and the subsequent kilonovae.
%
We summarize our main findings for both the ALV network and ET-B/C in Fig.~\ref{fig:summary}.
\begin{table}[h]
\centering
\begin{tabular}{lccc}
\hline\hline
 & ET-B & & ET-C\T\B\\
\hline
  $D_\text{BNS} $& 87\,Mpc& &{613\,Mpc}\T\\
  $R(D_\text{BNS}) $& $1^{+2}_{-1}\,\text{yr}^{-1}$&\hspace{5mm} &{$355^{+730}_{-280}\,\text{yr}^{-1}$}\T\\
  $\bar\rho_F(D_\text{BNS})$ & 420 &&{58}\T\B\\
\hline\hline
\end{tabular}
\caption{Horizon distances of ET-B and ET-C assuming $T_\text{AW} =1\,$hour. $R(D_\text{BNS})$ is the BNS merger rate within a volume of $D_\text{BNS}^3$
obtained by rescaling the rate inferred from Advanced LIGO's O1, O2 observing periods \cite{GW170817}. $\bar\rho_F(D_\text{BNS})$ is the total SNR accumulated due to a BNS inspiralling at $D_\text{BNS}$ [see Eq.~(\ref{eq:rhoF_ET})].}\label{table:horizon}
\end{table}
%
%
%
%
%
\section{Black hole neutron star inspirals}\label{sec:BH_NS}
In this section, we investigate the possibility of witnessing the tidal distruption of a neutron star by a black hole before the merger.
The treatment of the inspiral is the same as before, however the fate of the neutron star depends strongly on the system's intrinsic parameters
such as the black hole mass and spin, and the neutron star's equation of state (EOS) and compactness $\mathcal{C}_\text{NS}\equiv G M_\text{NS}/(R_\text{NS}c^2)$. 
The evolution of black hole-neutron star (BH-NS) binaries is a very active field of research at the interface of numerical relativity and high-energy astrophysics
(see Ref. \cite{Shibata:2011jka} for a comprehensive review) requiring fully general relativistic magnetohydrodynamic treatment including neutrino transport equations.
The complexity of evolving the dynamics of these systems require large computational resources and long runtimes.
As such, the details are beyond the scope of this article, but we can use a mostly Newtonian treatment to serve our purposes. 

A necessary, but insufficient, condition for tidal disruption (TD) is that
\be
\f{2 G M_\text{BH} (c_R R_\text{NS})}{r^3} \gtrsim \f{G M_\text{NS}}{(c_R R_\text{NS})^2} \label{eq:TD_cond1},
\ee
where $r$ is the orbital radius, $R_\text{NS}$ is the radius of the NS and $M$ denotes masses. 
$c_R R_\text{NS}$ is the semi-major axis of the elongated oblate spheroid that represents the tidally distorted shape of the NS to leading order with
$c_R >1 $. 
The above condition roughly implies
\be
\f{M_\text{BH}}{M_\text{NS}} \gtrsim \left(\f{r}{R_\text{NS}}\right)^3 \label{eq:TD_cond2}.
\ee
We wish to maximize the chances of observing a TD event, therefore we ingrain the BH-NS binary with certain desirable features
some of which can be deduced from Eqs.~(\ref{eq:TD_cond1}, \ref{eq:TD_cond2}). These are
\begin{enumerate}[(i)]
 \item Small separation. As the Newtonian tidal force due to the black hole is proportional to $M_\text{BH}/r^3$, $r$ has to be
 minimized to provide maximum tidal force. 
 However, for $r< r_\text{ISCO}$ the NS simply plunges into the BH; therefore the minium separation is given by the ISCO radius.
 \item Low black hole mass. In the test-mass case, $r_\text{ISCO} \propto M_\text{BH}$, so the maximum Newtonian tidal force roughly scales as $M_\text{BH}/r_\text{ISCO}^3\propto M_\text{BH}^{-2}$. 
 Hence, $M_\text{BH}$ must also be minimized. 
 There is a lower bound to it set from astrophysical observations and stellar evolution models, which currently yield $M_\text{BH} \gtrsim 5 M_\odot$ \cite{Farr:2010tu, Raithel:2017nlc, Ozel:2010su, Wiktorowicz:2013dua, Christian:2018mjv}.
 \item High black hole spins. $r_\text{ISCO}$ is much smaller in the case of a prograde NS orbit around a spinning Kerr black hole.
 For instance, for the maximum dimensionless BH spin value of $a\equiv c J_\text{BH}/(G M^2_\text{BH})=1$, $r_\text{ISCO}$ for prograde orbits is 1/6th of the Schwarzschild value $6GM_\text{BH}/c^2$ whereas for $a=1$ retrograde orbits, $r_\text{ISCO}$ is 3/2 of the Schwarzschild value.
 With smaller ISCO separation, there is much more tidal force exerted on the NS before it plunges in.
 \item Less compact neutron stars. Eq.~(\ref{eq:TD_cond1}) can be rearranged to isolate $\mathcal{C}_\text{NS}$
 on the right hand side. Smaller values for $\mathcal{C}_\text{NS}$ mean that TD is more likely to occur for $r>r_\text{ISCO}$.
\end{enumerate}

Although the results from each numerical study of BH-NS mergers may somewhat vary what is clear is that we must have high spin and low mass for the black hole for tidal disruption to happen. 
We are thus led to set $M_\text{BH} =5 M_\odot$. 
This raises an interesting astrophysical question as to whether it is possible to have such light-weight, high-spin black holes assuming that
they are spun up by accretion from their secondary (that will later on become a NS) in binary systems, thus they will have to gain mass.
It is possible that the mass gain pushes $M_\text{BH}$ too high to either cause a direct plunge of the NS before TD or decrease $\tau_\text{insp}$ considerably
to eliminate any chances of getting an advance warning.
However, if the spin of the BH comes from the supernova explosion mechanism as argued for G1915 in Ref.~\cite{McClintock:2006xd} then it is possible to have a 
low-mass, high-spin BH in a binary system\footnote{We thank Dr. Morgan Fraser of UCD Astrophysics
for pointing this out.}.

There is additional motivation to choose low BH mass as 
 $\tau_\text{insp} \propto M_c^{-5/3} = (1+Q)^{1/3}/Q$ for $Q\equiv M_\text{BH}/M_\text{NS} > 1$, which scales as $Q^{-2/3}$ as $Q\to\infty$ , i.e., binaries with heavier BHs inspiral faster
 so it is desirable to minimize $Q$.
 Setting $M_\text{NS}=1.4 M_\odot$ as before gives $Q=5/1.4  \simeq 3.571$. $\mathcal{C}_\text{NS}$ is determined by the radius of the NS constrained to be in the $9.9 - 11.2\,$km range\footnote{Recent work by the ALV collaboration \cite{GW170817_radii}) give bounds of $R_\text{NS}=9.1 -11.8\,$km, more or less consistent with astrophysical results.}\,\cite{Ozel:2016oaf}
 which results in $0.13\lesssim \mathcal{C}_\text{NS} \lesssim 0.15$ 
 for our chosen NS mass. 
If we pick the more pessimistic $\mathcal{C}_\text{NS} = 0.15$ bound then we must have $Q\lesssim 5$ and $a\gtrsim 0.5$ (prograde) for TD to occur \cite{Ferrari:2008nr}
and $Q< 3$ in the case of $a=0$ \cite{Kyutoku:2011vz}.

So, we simply assume that such lightweight, high-spin BHs exist in BH-NS binaries and compute strains and SNRs accordingly.
A quick calculation shows that these systems yield $T_\text{AW} \lesssim 30\,$seconds in the ALV network of 2020s. 
Therefore, we simply focus on the capabilities of ET which we summarize in Table~\ref{table:ET_BH_NS} below.
\begin{table}[h]
\centering
\begin{tabular}{lccccc}
\hline\hline
$D\,$(Mpc) & \multicolumn{2}{c}{ET-B} &  & \multicolumn{2}{c}{ET-C}\T\B\\
\hline
{}&  \ \hspace{6mm}$T_\text{AW}$ \hspace{8mm} & $\bar{\rho}_{F}$ &{}  & $\hspace{6mm}T_\text{AW} \hspace{10mm}$ & $\bar{\rho}_{F}$\T\B\\

100 &   28.2\,minutes & 503 &{\qquad} &  2.17\,hours & 601\T\\
200 & 8.21\,minutes & 251  &{\qquad} & 1.10\,hours & 300 \\
400 &  2.00\,minutes & 124 &{\qquad} & 34.2\,minutes & 148\\
1000 & 14.3\,seconds & 49.0 &{\qquad} & 13.9\,minutes& 58.5\\
\hline\hline
\end{tabular}
\caption{Same as Table~\ref{table:ET}, but now for a $5 M_\odot- 1.4 M_\odot$ black hole-neutron star binary.
The SNRs are even higher than before, however the advance warning times have decreased significantly.}\label{table:ET_BH_NS}
\end{table}

From the table, we can see that the range of ET-C for these binaries with $T_\text{AW} \approx 1\,$hour is roughly 200\,Mpc.
Given that the LIGO O1 merger rate for these systems is $ R < 3600\,$Gpc$^{-3}$yr$^{-1}$ \cite{Abbott:2016ymx},
there could be $\approx 30$ such mergers per year within ET-C's range of $\approx 200\,$Mpc. 
We must emphasize that this should be viewed
as an upper limit as low-mass, high-spin black holes such as ours above are expected to be rare.
Theoretical studies indicate that $\lesssim 10\%$ stellar-mass BHs have masses $\lesssim 5 M_\odot$ \cite{Fryer:1999ht}.
This seems to be supported at least by observations of galactic black holes \cite{Ozel:2010su}.
Therefore, we realistically expect ET-C to forecast $\lesssim 3$ tidal disruption events per year.
%
%

For the curious reader, we briefly repeat our computations for binary black hole (BBH) inspirals.
As heavier systems inspiral too fast, we limit ourselves
to $5M_\odot-5M_\odot$ BBHs and only consider ET-C.
For these systems, we find that $T_\text{AW} \gtrsim 1\,$hour only for $D\lesssim 190\,$Mpc, with $\bar{\rho}_F$ exceeding 500 due to the increased total mass. Even for a source at 100\,Mpc, we obtain $T_\text{AW}\approx 35\,$minutes at best.
As such lightweight black holes
are expected to be rare and given the current BBH merger rate of 12-213\,Gpc$^{-3}$yr$^{-1}$ \cite{Abbott:2017vtc}, which translates to $\ord(0.1) - \ord(1)$ BBH mergers yearly
within 190\,Mpc, we statistically expect no forecasts of these events. 
Nonetheless, ET and CE will detect hundreds to thousands of BBH mergers out to redshifts of $z> 10$.

\section{Outlook}
By constructing representative models of the near-future ground-based gravitational-wave interferometers, we investigated the possibility of
forecasting gamma-ray burts resulting from the merger of two neutron stars.
We showed that we do not expect the Advanced-LIGO-Virgo network to provide such a forecast in the 2020s unless we get extremely lucky
and detect a binary neutron star inspiral in our galaxy, which is roughly a one-in-a-trillion shot (in years). 
However, the odds completely change in our favour with Einstein Telescope's B and especially C configurations, 
the latter of which is expected to forecast $\gtrsim \ord(10^2)$ gamma-ray bursts per year as we show in Table~\ref{table:horizon}. 
The same configuration should also forecast up to three tidal disruption events per year in which a high-spin, low-stellar-mass
black hole tidally tears apart its companion neutron star.
The beginning of operation for the C configuration will roughly coincide with the launch of the LISA mission.
Additionally, there are proposals to launch another space interferometer called DECIGO which will operate
in the $0.1 - 10\,$Hz range thus bridging the gap between LISA and the ground-based interferometers \cite{Sato:2009zzb, Kawamura:2006zz}.
With the gravitational-wave sky virtually covered from $10^{-4}$ to $10^3\,$Hz, we will suffer from the 
embarassment of the riches in the 2030s. In short, the future of gravitational and multi-messenger astronomy is bright.

\acknowledgements 
I acknowledge support by the EU H2020 under ERC Starting Grant, no. BinGraSp-714626.
I am additionally grateful to Conor O'Toole for endless feedback and Niels Warburton, Richard A. Matzner, and Alberto Sesana for reviewing this manuscript.

\bibliographystyle{apsrev4-1}
\bibliography{references}

\end{document}